\documentclass[sigconf]{acmart}

\AtBeginDocument{%
  \providecommand\BibTeX{{%
    \normalfont B\kern-0.5em{\scshape i\kern-0.25em b}\kern-0.8em\TeX}}}

\acmISBN{979-8-4007-0686-8/24/10}
\acmDOI{10.1145/3664647.3681021}

\usepackage{arydshln}
\usepackage{booktabs}
\usepackage{colortbl}
\usepackage{xcolor}
\usepackage{multirow}
\usepackage {verbatim}
\usepackage{graphicx}
\usepackage{caption}
\usepackage{hyperref}
\usepackage{float}
\usepackage{nextpage}
\usepackage{lipsum}
\usepackage{footmisc}
\usepackage{blindtext}

\begin{document}

\title{Semantic-aware Representation Learning for Homography Estimation}

\author{Yuhan Liu$^{1,*}$, Qianxin Huang$^{1,2,*}$, Siqi Hui$^{1}$, Jingwen Fu$^{1}$, Sanping Zhou$^{1}$,  Kangyi Wu$^{1}$, Pengna Li$^{1}$, Jinjun Wang$^{1,\dagger}$}

        \affiliation{
         \institution{
            $^1$ National Key Laboratory of Human-Machine Hybrid Augmented Intelligence,\\
             National Engineering Research Center for Visual Information and Applications,\\
             and Institute of Artificial Intelligence and Robotics, Xi’an Jiaotong University\\
            $^2$ Huawei Inc
        }
        \city{}
        \country{}
        }
        \email{liuyuhan200095@stu.xjtu.edu.cn}
	\renewcommand{\shortauthors}{Yuhan Liu, et al.}

\begin{abstract}

Homography estimation is the task of determining the transformation from an image pair. Our approach focuses on employing detector-free feature matching methods to address this issue.  Previous work has underscored the importance of incorporating semantic information, however there still lacks an efficient way to utilize semantic information. Previous methods suffer from treating the semantics as a pre-processing, causing the utilization of semantics overly coarse-grained and lack adaptability when dealing with different tasks. In our work, we seek another way to use the semantic information, that is semantic-aware feature representation learning framework.
Based on this, we propose SRMatcher, a new detector-free feature matching method, which encourages the network to learn integrated semantic feature representation.
Specifically, to capture precise and rich semantics, we leverage the capabilities of recently popularized vision foundation models (VFMs) trained on extensive datasets.  Then, a cross-images Semantic-aware Fusion Block (SFB) is proposed to integrate its fine-grained semantic features into the feature representation space. In this way, by reducing errors stemming from semantic inconsistencies in matching pairs, our proposed SRMatcher is able to deliver more accurate and realistic outcomes. 
Extensive experiments show that SRMatcher surpasses solid baselines and attains SOTA results on multiple real-world datasets. Compared to the previous SOTA approach GeoFormer, SRMatcher increases the area under the cumulative curve (AUC) by about 11\% on HPatches. 
Additionally, the SRMatcher could serve as a plug-and-play framework for other matching methods like LoFTR, yielding substantial precision improvement. Code is available at \url{https://github.com/lyh200095/SRMatcher}.

\end{abstract}



\maketitle

\let\thefootnote\relax\footnotetext{$^{*}$Contributed equally.}

\let\thefootnote\relax\footnotetext{$\dagger$Corresponding author.}

\begin{figure}[t]
\centering
\includegraphics[width=\linewidth]{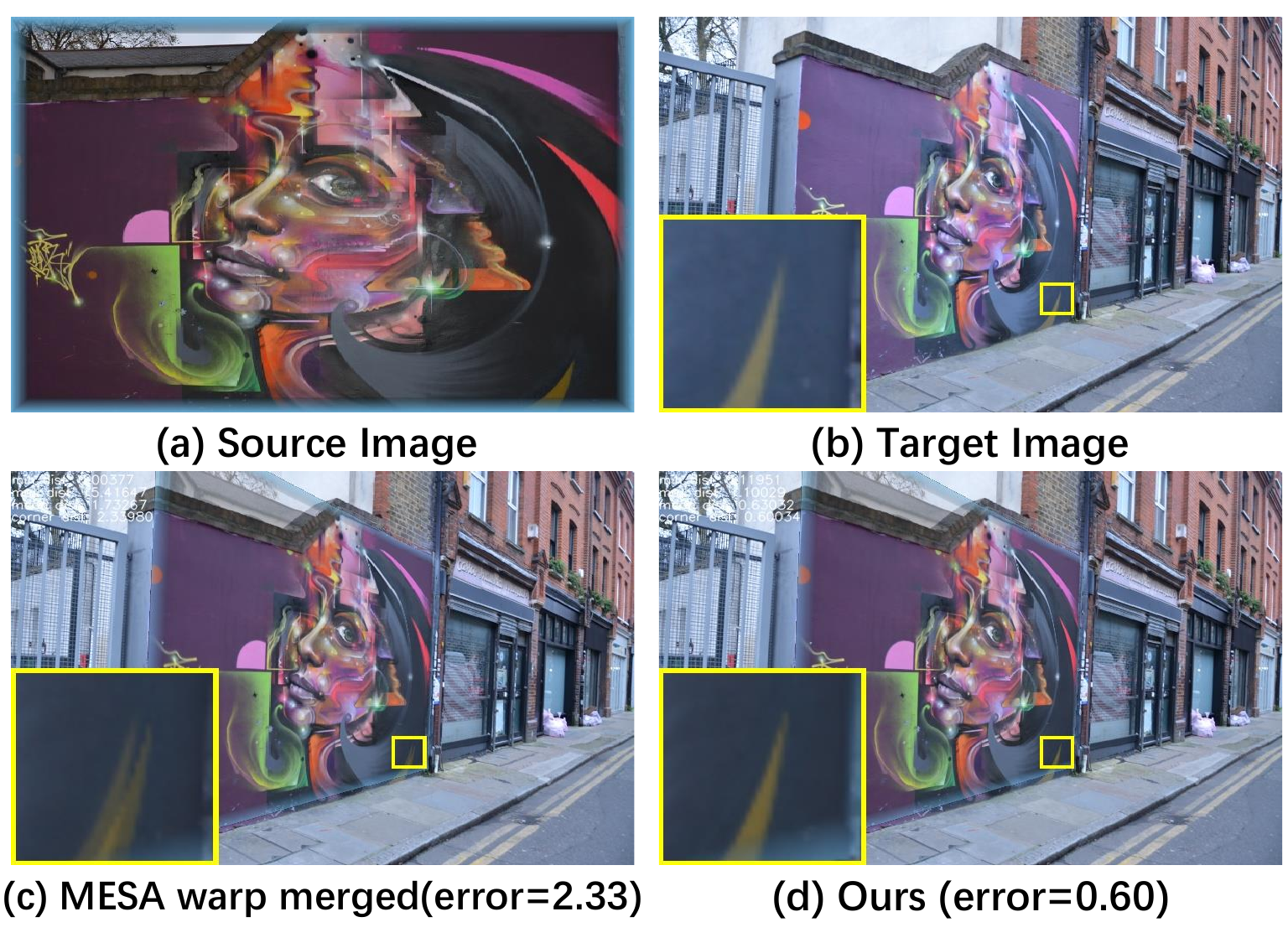}
\captionsetup{skip=3pt}
\caption{Homography transformation results by our proposed SRMatcher and MESA \cite{zhang2024mesa}. The blue box line was added artificially to highlight the range of homologous transformations. The yellow boxes show the real scenes cropped in the target image. (c) and (d) are generated by superimposing the warped source image on the target image, showing our SRMatcher acquiring more accurate and realistic outcomes.}
\label{fig_first}
\vspace{-1.5em}
\end{figure}

\section{Introduction}
Homography estimation, often referred to as perspective transformation or planar projection from source image to target image as Figure \ref{fig_first} shown, is a core issue in the fields of computer vision and robotics \cite{hong2022unsupervised, zhang2020content}. This process entails calculating a matrix to align points between two images captured from distinct viewpoints. This estimation is pivotal for numerous applications, including including image/video stitching \cite{guo2016joint,zaragoza2013projective}, camera calibration \cite{zhang2000flexible}, HDR imaging \cite{gelfand2010multi}, and SLAM \cite{mur2015orb,zou2012coslam}. Feature matching techniques are widely used to establish feature correspondences in tasks related to homography estimation. 
The traditional matching methods \cite{lowe2004distinctive, rublee2011orb} rely on the quality of keypoints, and can sometimes falter in challenging scenarios such as low texture and blurred \cite{sun2021loftr}. A novel detector-free matching methods have surfaced as more dependable alternatives.  
These techniques generate dense feature maps without the need for feature detection and execute dense matching on a pixel-wise basis \cite{chen2022aspanformer, jiang2021cotr, tang2022quadtree, cao2023improving, liu2023geometrized}. Existing methods typically extract local features related to position and context but overlook the semantic information of matching pairs, which is crucial for feature matching.

Previous work \cite{zhang2023searching, zhang2024mesa} such as MESA do consider semantics, these methods view the matching issue as a search problem, searching corresponding matching points for specific source points in the target image. The introduction of semantics aims to narrow the search space, by establishing area matches the initial search space for point matching is confined to the matched area. This method resembles the human cognitive process, humans first identify matched areas using semantic information and subsequently search for corresponding points within those areas. However, their work faces two notable problems: \textbf{1) The first problem} is these methods restrict the application of semantic priors to coarse-grained scenarios, failing to fully leverage the knowledge that semantic extraction networks can offer. The goal of incorporating semantic priors into the matching network is to reduce erroneous matches by constraining the semantic consistency of the matching results. However these area matching method merely narrows down the matching area, and semantics can not guide the matching result in subsequent steps, failing to achieve the initial goal. \textbf{2) For the second problem}, the interaction between semantic information and matching is designed for particular approaches. When dealing with different tasks or data types, processing methods such as area size and quantity need optimization for particular applications. This means these patch-wise method suffer from a lack of adaptability and generalize issues, limiting the ability of the algorithm to understand the overall scene.

We identify the key driver of these two problems as the ill-use of the semantic information, specifically, treating semantics as merely a pre-processing step. These methods initially identify semantic area matches based on semantic segmentation and then acquire precise correspondences within these regions using a standard point matcher. This process is intuitive and closely resembles human thought processes \cite{zhang2023searching, truong2023topicfm+}. However, for the human cognitive system, an implicit prerequisite in the task of feature matching is that the features being matched should share the same semantics \cite{giang2023topicfm, wolfe2011visual, wu2014guidance}. Based on this, we propose another way to use the semantic information that encourages the network to learn semantic-aware representation.

In this paper, we propose SRMatcher, the first semantic-aware representation learning framework which consists of two essential parts: 
\textbf{(a)} Semantic Extractor. \textbf{(b)} Semantic-aware Fusion Block(SFB). The key challenge to achieving the semantic-aware representation is\textit{ how to generate similar representations for semantically similar points in image pairs.} \textbf{For (a)}, we propose to use the DINOv2 \cite{oquab2023dinov2} as the semantic extractor to acquire richer and more diverse semantic information.  This considers the outstanding performance of vision foundation models (VFMs) in tasks related to semantics.
\textbf{For (b)}, we develop a Semantic-aware Fusion Block (SFB) to utilize fine-grained semantic features to contribute to matching quality. Different from previous works \cite{wu2023learning, li2024sed, wu2023seesr}, SFB executes cross-images feature fusion which means image features should not only integrate their own semantic information but also consider the semantics across images. This difference comes from considering the nature of the task, that is the image feature should be enhanced with semantic information of other images to find semantically identical points. In more detail, we employ a semantic-guide interactions block (SGIB) to establish the interactions between image features and semantic features within the SFB.

In summary, this work makes the following contributions:

\begin{itemize}
    \item We explore a new way to integrate semantics into the feature matching model for homography estimation. For the first time, we propose SRMatcher, a novel detector-free feature matching method by integrating semantics from vision foundation models (VFMs) into the network.
\end{itemize}
\begin{itemize}
    \item To enhance the semantic guidance for the matching model, we introduce the Semantic-aware Fusion Block (SFB) to conduct cross-images semantic feature fusion, whereby features are prompted to concurrently consider the semantics from multiple images.
\end{itemize}

\begin{itemize}
    \item Extensive experiments demonstrate that SRMatcher achieves the SOTA performance on homography estimation tasks and other downstream vision tasks. Additionally, our SRMatcher can be seamlessly incorporated into various benchmarks for detector-free feature matching methods in a plug-and-play fashion.
\end{itemize}

\section{RELATED WORK}

\subsection{Local Feature Matching}
Local feature matching can be categorized into detector-based and detector-free methods. Detector-based methods operate in four stages: detecting keypoints, extracting local features for each keypoint, matching features based on content, and finally fitting a homography using the identified matches. SIFT \cite{lowe2004distinctive} features exhibit invariance to rotation and scale. With the advent of deep learning, many learning-based methods have been proposed, to further enhance the robustness of descriptors under varying illumination and viewpoint changes, such as Superpoint \cite{detone2018superpoint}, D2-Net \cite{revaud2019r2d2}. SuperGlue \cite{sarlin2020superglue} utilizes Transformer \cite{vaswani2017attention} to establish correspondences between two sets of local features generated by SuperPoint. The drawback of detector-based methods lies in their excessive reliance on the quality of keypoint detection, which is particularly unsatisfactory in low-texture and repetitive texture regions.
Detector-free methods establish dense matching relationships directly between pixels, eliminating the need for keypoint detection. LoFTR \cite{sun2021loftr} utilizes self- and cross-attention mechanisms on feature maps derived from CNN, producing matches progressively from coarse to fine detail. GeoFormer \cite{liu2023geometrized} using the RANSAC algorithm for attention region search, the computation scope of attention is confined to a specific region, enabling the use of a standard transformer for processing. 
While these approaches are methodologically robust, they fall short in capturing high-level contexts such as semantic information. They fail to explicitly represent the semantic correspondences within image pairs and lack clarity in their interpretability. Conversely, our method, through a Semantic-aware Fusion Block(SFB), can utilize fine-grained semantic-aware features, making our matching outcomes significantly more interpretable and accurate.

\subsection{Semantics Feature Matching}
Lately, methods guided by semantics have demonstrated the dependability of semantic priors. SGAM \cite{zhang2023searching} incorporating semantics segmentation maps as an additional input, region matching based on identical semantics is conducted within the image before point matching. However, SGAM utilizing the direct semantic label results in the area matching's performance being highly sensitive to the precision of semantic labeling. Conversely, MESA \cite{zhang2024mesa}  employs pure image segmentation for area matching, offering a more pragmatic approach and overcoming the limitations associated with direct semantics labels. Nevertheless, the semantic representation of this approach is coarse-grained, and semantic inconsistency still exists in the matched regions, when it comes to detailed point matching. In comparison, we explore the adaptive latest network frameworks that use a novel interplay between semantic priors and the original tasks, allowing the model to utilize rich semantic priors from a pre-trained encoder. Our approach integrates image appearance features with high-dimensional semantics, integrate fine-grained  semantic into feature representation space. This allows our method to be a fully handcrafted, plug-and-play framework.

\subsection{Vision Foundation Models}
Leveraging extensive pre-training, foundational models in vision have garnered significant achievements in the field of computer vision, shows excellent performance in various downstream tasks \cite{liu2023matcher, zhang2024mesa, li2024sed, edstedt2023roma}. Inspired by the masked language modeling \cite{devlin2018bert} from natural language processing, Masked Autoencoder (MAE) \cite{he2022masked} utilizes an asymmetric encoder-decoder architecture for masked image modeling, facilitating the effective and efficient training of scalable vision transformer models, with MAE displaying outstanding fine-tuning capabilities in a variety of downstream tasks. CLIP \cite{radford2021learning} acquires image representations from the ground up using 400 million image-text pairs, showcasing remarkable zero-shot image classification skills. Through discriminative self-supervised learning at both the image and patch levels, DINOv2 \cite{oquab2023dinov2} acquires versatile visual features applicable to a broad spectrum of downstream tasks, also exhibiting notable patch-matching proficiency that identifies semantic components executing similar functions across diverse objects or species. Motivated by these developments, our objective is to utilize more fine-grained semantics from DINOv2 to direct our SRMatcher.

\section{APPROACH}

\subsection{Overall Framework}
\label{section:3.1}
\begin{figure*}
  \includegraphics[width=\textwidth]{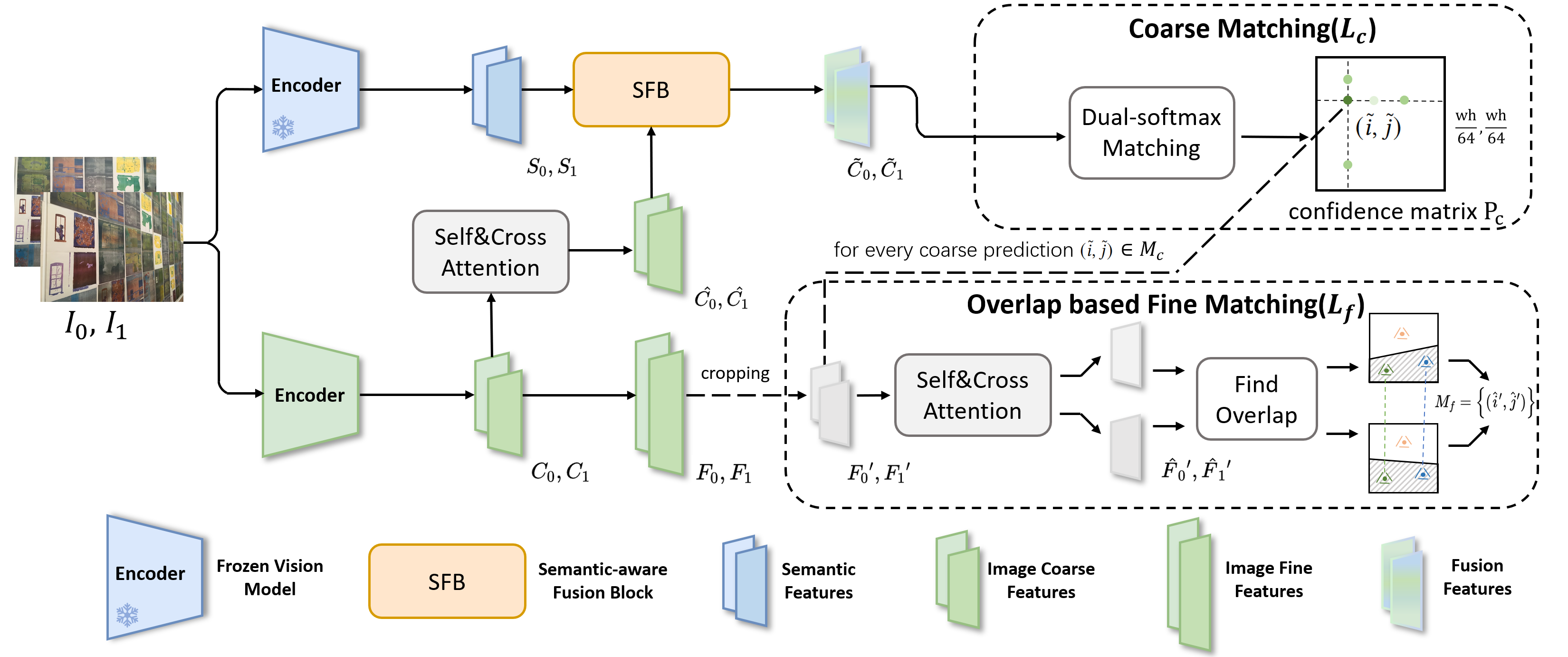}
  \captionsetup{skip=-1pt}
  \caption{Overview of our SRMatcher for detector-free local feature matching and followed by homography estimation. With a pretrained Semantic extract network, our SRMatcher utilizes fine-grained features to improve the matching results. The SFB enable interactions between image features \(\hat{C_{0}}\) and \(\hat{C_{1}}\) and semantic features \(S_{0}\), \(S_{1}\), produce the fusion features. The coarse matching block generates pixel-to-pixel matches \(M_{c}\) at 1/8 scale. Subsequently, the \(M_{c}\) input into the overlap-based fine matching to yield fine matches \(M_{f}\) at 1/2 scale. }
 
  \label{fig_2m}
  \vspace{-5mm}
\end{figure*}

As our approach is built upon LoFTR, we initially provide a concise description of this method. Given a pair of input images $I_0$ and $I_1$ with dimensions $w \times h$ in gray-scale, a 2D-CNN (ResNetFPN \cite{lin2017feature}) extracts two feature maps per image: a coarse-level feature map at $1/8$ of the original size and a fine-level feature map at $1/2$ of the original size. The coarse features \(C_{0}\) and \(C_{1}\) are input into a attention mechanisms to produce \(\hat{C_{0} }\) and \(\hat{C_{1} }\). Then the coarse matching module generates a pixel-to-pixel confidence matrix $P_c$. 
Coarse-level matches $M_c = \{(\tilde{i}, \tilde{j})\}$ are established by applying a threshold to $P_c$ and conducting a mutual nearest neighbor (MNN) search.
Given $M_c$ and the fine-level features \(F_{0}\) and \(F_{1}\), the fine-matching module computes sub-pixel matches in the following way. For each pair $(\tilde{i}, \tilde{j})\in M_c$, a square region of dimensions $w \times w$ centered around $\hat{i}=\tilde{i}\times4$ is extracted from \(F_{0}\), and a similar region centered around $\hat{j}=\tilde{j}\times4$ is extracted from \(F_{1}\). The optimal correspondence for pixel $\hat{i}$ within the \(F_{1}\) region determines the set of sub-pixel matches $M_f$. Inherited from LoFTR, we propose SRMatcher the semantic-aware feature representation learning framework. The comprehensive structure of our SRMatcher is shown in Figure \ref{fig_2m}. 

Given a pair of images, we initially process them through a CNN backbone to extract coarse and fine image features. And a frozen pre-trained semantic extractor from a large vision model to extract the semantics. Subsequently, the enhanced image features by self- and cross-attention undergo processing in our SFB module to integrate with fine-grained semantic features. Following this, coarse matching is utilized to establish patch-level matches. Ultimately, the overlap based fine-matching is employed to predict fine-level matches. With the semantically informed features, the network can perform semantic-aware feature matching with our proposed framework. The problem definition of the semantic-aware representation learning framework is as follows:

Given a pair of image $I_0$ and $I_1$\(\in \mathbb{R}^{H \times W \times 3}\) with width  $W$ and height $H$. Combining with fine-grained semantic features, the matching process can be modeled as follow, first:
\begin{equation}
  S_{i}={F} _{semantic}(I_{i};\theta _{s} ), i=0,1
\end{equation}
where $S_{i}$ represents the fine-grained semantic features. \({F}_{semantic}\) denotes the pre-trained semantic extraction network, with \(\theta _{s}\) being frozen during the training stage. $S_{i}$ is then utilized as input:
\begin{equation}
  M_{h} ={F} _{match}(I_{i},S_{i};\theta _{m} )
\end{equation}
where $M_{h}$ is the matched result and \({F}_{match}\) represents the matching network. Throughout the training stage, \(\theta_m\) is updated by minimizing the objective function with the collaboration of \(S_{i}\), while \(\theta_s\) is frozen:
\begin{equation}
  \hat{\theta _{m} } ={argmin}{L} (\hat{M_{h} },{M_{h} },S_{i})
\end{equation}
where $\hat{M_{h}}$ is the ground truth, ${{L}} (\hat{M_{h} },{M_{h} },S_{i})$ is the objective function of semantic-aware matching model. The details of our semantic-aware feature representation learning framework will be elaborated in the following section.

\subsection{Semantic-aware Fusion Block}
\label{section:3.2}

\begin{figure*}
  \includegraphics[width=\textwidth]{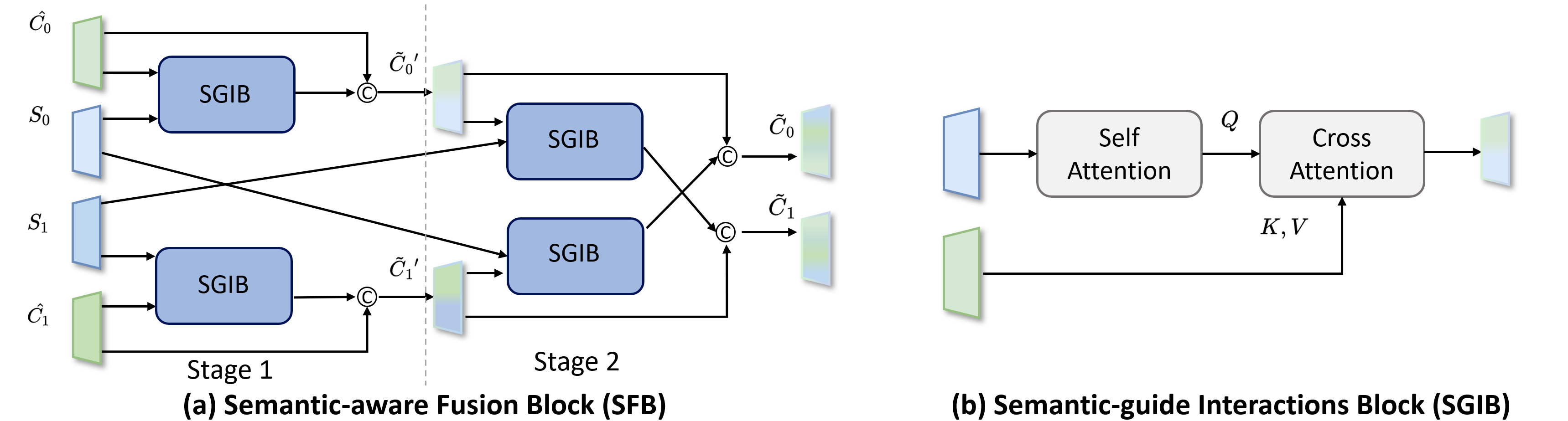}
  \captionsetup{skip=-1pt}
  \caption{Architecture of the (a)semantic-aware fusion block(SFB) and (b)semantic-guide interactions block(SGIB). The SFB fuses image features and semantic features across images. Inside SFB, SGIB computes cross-attention that the image features as key K / value V and the semantic feature as query Q. }
  \label{fig_3m}
  \vspace{-5mm}
\end{figure*}

When enhancing image features using semantic features, it's crucial to address the discrepancies between the two sources. To mitigate this, we introduce the SFB module to refine the image feature maps, as depicted in Figure \ref{fig_3m} (a). Serving as a bridges between the semantic net and the matching net, the SFB modules create connections between these two distinct tasks.

The efficacy of semantic extraction is crucial for the semantic-aware feature representation learning framework. The motivation is that fine-grained semantics will enable the network to generate more precise matching results. The advanced vision foundation models (VFMs), renowned for their robust representational prowess, are ideal for this task. Particularly, the DINOv2 model stands out for its effectiveness and versatility across diverse applications \cite{yang2024depth, liu2023matcher}, making it a preferred choice for semantic extraction. For our purposes, we select the pretrained DINOv2 model as our semantic extractor to obtain pixel-level semantic information. 

Upon acquiring the semantics \(S_{i}\), we utilize the proposed Seman-tic-aware Fusion Block (SFB) to integrate semantic features into the matching model. With the enhanced image features of coarse-level \(\hat{C_{0} }\) and \(\hat{C_{1} }\), our goal is to warp the semantic features from the images to the matching model. Within the SFB, we have designed a cross-images feature fusion strategy, as image features are to be augmented not only based on their own semantic information but also taking into account the semantic information from across different images. For the input features \(S_{i}\) and \(\hat{C_{i} }\), we propose a semantic-guide interactions block (SGIB) within SFB to more effectively integrate fine-grained semantics features with image features.\\
\textbf{Semantic-guide Interactions Block (SGIB). }
In the first stage of SFB, the image feature and semantic features from the same image are fused.
The SGIB module conducts a pixel-level interaction between image features \(\hat{C_{i} }\) and semantic features \(S_{i}\) and output the refined feature map \(\tilde{C{} _{i}{}  }{}'\). As shown in Figure \ref{fig_3m} (b), given the semantic features \(S_{i}\), first go through a self-attention(SA) module and then supplied to the cross-attention module as query \(Q_{i}\), then the image features \(\hat{C_{i} }\) are identified as the keys \(K_{i}\) and values \(V_{i}\):
\begin{equation}
\begin{gathered}
    Q_{i}=\mathcal{SA} (S_{i} )\\
    \tilde{C{} _{i}{}  }{}'=Concat(\hat{C_{i}},(Softmax(Q_{i}K_{i}^{T}/\sqrt{C}))V_{i})
\end{gathered} 
\end{equation}
where C denote the channel and the Concat represent the concatenate operation. And then at the second stage of SFB, the image feature and semantic features from the different image are fused. The calculation process is similar to the first stage, the keys \(K_{i}{}'\) and values \(V_{i}{}'\) are produced by \(\tilde{C{} _{i}{}  }{}'\):
\begin{equation}
\begin{gathered}
    {Q_{j}}{}'=\mathcal{SA} (S_{j} ), j=0,1\\
    \tilde{C{} _{i}{}  }=Concat(\tilde{C{} _{i}{}  }{}',(Softmax(Q_{j}{}'K_{i}{}'^{T}/\sqrt{C}))V_{i}{}'), j\ne i
\end{gathered} 
\end{equation}
\textbf{Coarse Matching.}
Given the fusion features  \(\tilde{C{} _{i}{}}\), we adopt the LoFTR approach to generate coarse matches \(M_c\). We apply a dual-softmax function to process a confidence matrix \(P_{c}\), where \(P_{c}(i,j) \) signifies the probability of pixel $i$ and pixel $j$ is a correct match. This probability takes into account both low-level image information and high-level semantic information. By applying the threshold-based filter and the mutual nearest neighbor (MNN) search on \(P_c\) derived \(M_c\).

\subsection{Overlap based Fine Matching}
As mentioned in Section \ref{section:3.1}, we also integrate the patch-wise refinement module from LoFTR for sub-pixel matching accuracy enhancement. Retrospect that the fine matching process utilized by LoFTR is oversimplified: For every coarse matching $(\tilde{i}, \tilde{j})$, LoFTR finds its position $(\hat{i} , \hat{j})$ on fine-level feature map $F_{i}$ and then crop two local windows based the location. After being transformed by the LoFTR module, for each local window only the center $\hat{i}$ of the enhanced feature is used to find its matched pixel ${\hat{j}}' $. When $\hat{i}$ is not within the overlap of two windows, this method may result in matching errors. As a supplement, we propose an overlap based fine-matching method. After the windows cropping, we calculate pixel-to-pixel similarities between the two windows. After applying threshold-based masking, a mutual nearest neighbor(MNN) search is conducted to refine the matches in the overlap area of windows, we represent the predictions for fine-level matches \(M_f\) as:
\begin{equation}
M_{f}=\left \{(\hat{i}{}' ,  \hat{j}{}' )\mid \hat{i}{}' =\mathcal{W}_{1\to 0} (\hat{j}{}' )\wedge\hat{j}{}' =\mathcal{W}_{0\to 1} (\hat{i}{}' )    \right \}   
\end{equation}
where $\mathcal{W}$ is the warping operation, $\wedge$ is the and operation.

\subsection{Self-supervised Training}
Inspired by LoFTR, SRMatcher is trained using a self-supervised approach, eliminating the need for manual annotations. We create a geometrically transformed image \(I_1\) from an original image \(I_0\) by applying a predefined homography to \(I_0\). Utilizing this homography, the true correspondence \(j\) in \(I_1\) for each element \(i\) in \(M_f\) can be precisely determined. Consequently, a set of ground-truth matches \(G_f\) corresponding to \(M_f\) is dynamically produced. Similarly, we generate ground-truth matches \(G_c\) for \(M_c\). Data augmentation includes random sampling of homographies. We also introduce various random photometric distortions to the image pairs, such as adjustments in brightness and contrast, motion blur, and Gaussian noise.

As shown in Figure \ref{fig_2m}, the coarse matching module, and the fine matching module each generate two pixel-to-pixel similarity matrices: \(P_c\), and \(P_f\), respectively. Ideally, these matrices should closely align with their respective ground truths. Following, the Focal binary cross-entropy Loss (FL) is employed to optimize multi-scale matching, with the loss defined as:
\begin{equation}
\begin{gathered}
  L_{c}= -\frac{1}{\left | G_{c}  \right | }  {\textstyle \sum_{(i,j)\in G_{c} }^{}}log(P_{c}(i,j) )\\
  L_{f}= -\frac{1}{\left | G_{f}  \right | }  {\textstyle \sum_{(i,j)\in G_{f} }^{}}log(P_{f}(i,j) ) 
\end{gathered}  
\end{equation}

The final loss is composed of the losses for the coarse-level and the fine-level, the cumulative loss function is expressed as: \(L_{toatl}= L_{c}+L_{f}\).

\section{EXPERIMENTS}
\subsection{Implementation Details}
\textbf{Experimental data.} SRMatcher is trained on the outdoor Oxford-Paris dataset, adhering to the protocols established following \cite{liu2023geometrized}. This dataset is merged by the Oxford5K \cite{philbin2007object} and Paris6K \cite{philbin2008lost} datasets, collectively referred to as Oxford-Paris, and encompasses a variety of outdoor and urban scenes. For evaluation, SRMatcher is tested on diverse image datasets, including HPatches \cite{balntas2017hpatches}, ISC-HE \cite{liu2023geometrized} and Megadepth \cite{li2018megadepth}. These test datasets feature a range of image categories, from natural scenes with pronounced viewpoint and illumination shifts to extensively edited photographs. The training dataset's registered image pairs are generated automatically, ensuring the entire network training through a completely self-supervised approach.

\textbf{Implementation.} SRMatcher is developed in PyTorch. Given our computational resources (8 NVIDIA GeForce RTX 3090 GPUs), we set the larger image dimension to 640 for natural images during training. We use the Adam optimizer, with \(\beta\) values of (0.9, 0.999) and a starting learning rate of 0.001. Each mini-batch consists of a pair of images. Training is capped at 15 epochs. SRMatcher is adapted for LoFTR \cite{sun2021loftr} and its variations, including ASpanFormer \cite{chen2022aspanformer} and GeoFormer \cite{liu2023geometrized}, resulting in the creation of SRMatcher\_LoFTR, SRMatcher\_ASpan, and SRMatcher\_GeoFormer, respectively. The distinguishing feature among these three versions lies in the attention mechanism employed within the feature interaction module, specifically linear attention \cite{katharopoulos2020transformers}, span attention \cite{chen2022aspanformer} and geometrized attention \cite{liu2023geometrized}. For the inference phase, with the obtained fine matches \(M_f\), we employ \texttt{cv2.findHomography} with RANSAC for reliable homography determination.

\subsection{Evaluation on Natural Images}
\label{section:4.2}
\textbf{Test Data.} The extensively utilized HPatches \cite{balntas2017hpatches} dataset is employed, comprising 57 sequences with substantial illumination shifts and 59 sequences exhibiting significant viewpoint changes, making it a particularly demanding benchmark for homography estimation. 

\textbf{Metrics.} Following \cite{detone2018superpoint,sun2021loftr}, we employ corner correctness to assess the accuracy of the estimated homography. The four corners from the first reference image are mapped to the second image using the estimated homography. The area under the cumulative curve (AUC) of the corner error is reported for threshold values of 1, 3, 5, and 10 pixels, respectively. For evaluation, all test images are resized to have their shorter dimension equal to 480. 

\textbf{Comparative methods.} Matching approaches are categorized into two kinds of methods, i.e. detector-based feature matching and detector-free feature matching. We compiling a selection of 11 baseline methods as detailed below: 1) Detector-based feature matching methods including SIFT \cite{lowe2004distinctive}+RooTSIFT \cite{arandjelovic2012three}, Superpoint \cite{detone2018superpoint}, SuperGlue \cite{sarlin2020superglue} and R2D2 \cite{revaud2019r2d2}; 2) Detector-free feature matching methods including TopicFM \cite{giang2023topicfm}, AdaMatcher \cite{huang2023adaptive}, CasMTR \cite{cao2023improving}, MESA \cite{zhang2024mesa}, LoFTR \cite{sun2021loftr}, ASpanFormer \cite{chen2022aspanformer}, GeoFormer \cite{liu2023geometrized}. 
For LoFTR, ASpanFormer, GeoFormer, we test them with their original version and their upgraded version.
For fair comparison we retrained all these detector-free models using the same Oxford-Paris dataset. Due to the limitation of computing resources, we choose the 4c version of CasMTR with NMS. 

\begin{table}[t]
\caption{ Homography estimation results on HPatches. The \textbf{best} and \underline{second} results are highlighted.}\label{tab:PAUC_SN}
\centering
\vspace{-3mm}
\resizebox{\linewidth}{!}
{
\begin{tabular}{llllllll}
\toprule
\multicolumn{3}{l}{\multirow{3}{*}{Method}} & \multicolumn{4}{c}{Homography est. AUC} \\ \cmidrule(l){4-7} 
\multicolumn{3}{c}{}               & @1px$\uparrow$         & @3px$\uparrow$       & @5px$\uparrow$        & @10px$\uparrow$        \\ \midrule
\multicolumn{7}{l}{$Detector$-$based$ $matching:$} \\
\multicolumn{3}{l}
{SIFT~\cite{lowe2004distinctive}}      &-         & 46.3      & 57.4      & {70.3}     \\
\multicolumn{3}{l}
{Superpoint~\cite{detone2018superpoint}~\tiny{CVPRW'18}}      &-         & 43.4      & 57.6      & {72.7}     \\
\multicolumn{3}{l}
{R2D2~\cite{revaud2019r2d2}~\tiny{NIPS'19}}      &-         & 50.6      & 63.9      & {76.8}     \\
\multicolumn{3}{l}
{SuperGlue~\cite{sarlin2020superglue}~\tiny{CVPR'20}}      &-         & 53.9      & 68.3      & {81.7}     \\

\midrule
\multicolumn{7}{l}{$Detector$-$free$ $matching:$} \\

\multicolumn{3}{l}{TopicFM~\cite{giang2023topicfm}~\tiny{AAAI'23}}       &40.5            & 63.2       & 73.0       & 82.9      \\
\multicolumn{3}{l}{AdaMatcher~\cite{huang2023adaptive}~\tiny{CVPR'23}}       &41.1            & 64.2       & 73.8       & 83.3       \\
\multicolumn{3}{l}{CasMTR~\cite{cao2023improving}~\tiny{ICCV'23}}    &41.6    & {65.9}      & {{74.7}}      & {83.8}       \\
\multicolumn{3}{l}{MESA~\cite{zhang2024mesa}~\tiny{CVPR'24}}    &43.9    & {67.8}      & \underline{{76.8}}      & \underline{85.5}       \\

\midrule
\multicolumn{3}{l}{LoFTR~\cite{sun2021loftr}~\tiny{CVPR'21}}    &34.2               & 58.5       & 69.8       & 81.1       \\

\rowcolor[rgb]{.9,.95,.98}\multicolumn{3}{l}{SRMatcher\_LoFTR}   &38.6$_{+12.87\%}$     & 62.3$_{+6.49\%}$       & 72.1$_{+3.29\%}$       & 82.3$_{+1.48\%}$       \\ \midrule
\multicolumn{3}{l}{ASpan~\cite{chen2022aspanformer}~\tiny{ECCV'22}}     &36.1              & 59.9       & 71.1       & 81.6       \\

\rowcolor[rgb]{.9,.95,.98}\multicolumn{3}{l}{SRMatcher\_ASpan}   &40.9$_{+13.29\%}$     & 63.9$_{+6.68\%}$       & 74.1$_{+4.21\%}$       & 82.8$_{+1.47\%}$       \\ \midrule
\multicolumn{3}{l}{GeoFormer~\cite{liu2023geometrized}~\tiny{ICCV'23}}    &\underline{44.3}               & \underline{68.0}       & \underline{76.8}       & {85.4}       \\

\rowcolor[rgb]{.9,.95,.98}\multicolumn{3}{l}{SRMatcher\_GeoFormer}   &\textbf{49.2}$_{+11.06\%}$     & \textbf{71.2}$_{+4.71\%}$       & \textbf{79.3}$_{+3.26\%}$       & \textbf{87.0}$_{+1.87\%}$       \\ 
\bottomrule
\end{tabular}

}\label{tab:t1m}
\vspace{-8mm}
\end{table}

\begin{figure*}
  \includegraphics[width=\textwidth]{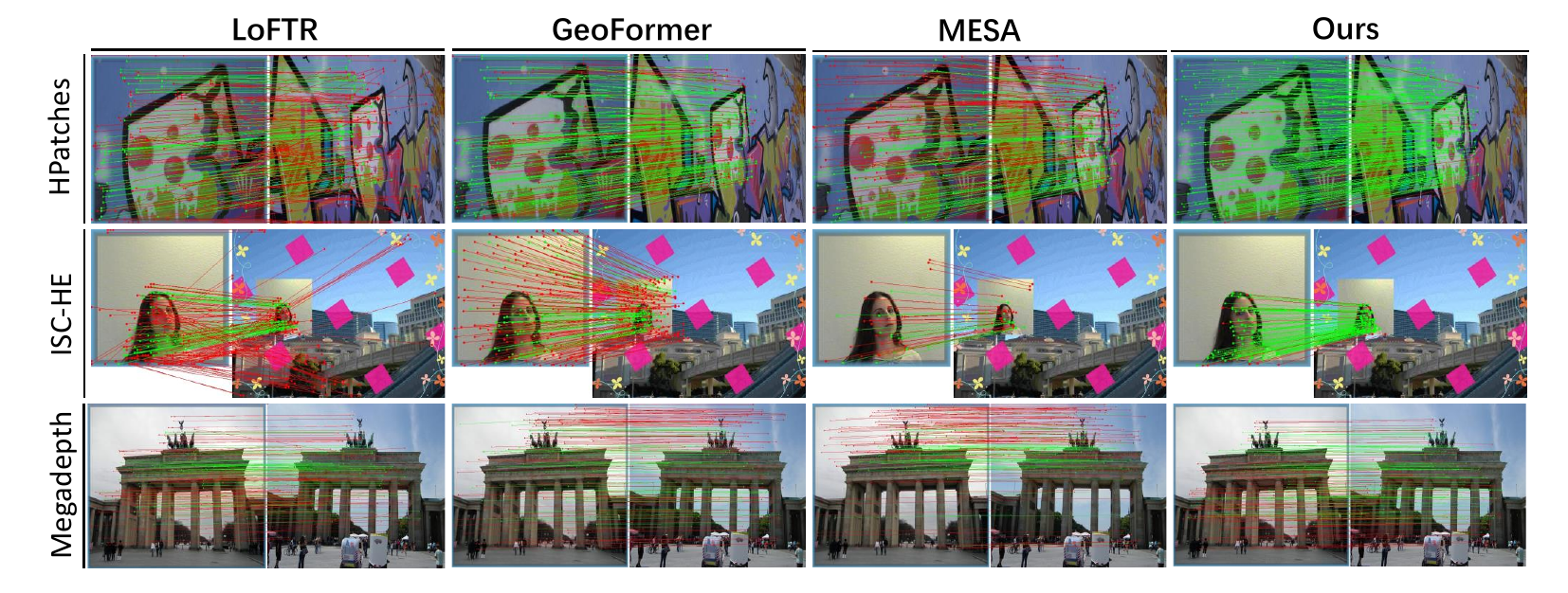}
  \captionsetup{skip=0.5pt}
  \caption{Qualitative of matching results with LoFTR \cite{sun2021loftr}, GeoFormer \cite{liu2023geometrized}, MESA \cite{zhang2024mesa}, and our SRMatcher. Points classified as inliers by RANSAC are displayed in green, while outliers are shown in red. }
  \label{fig_4m}
  \vspace{-4mm}
\end{figure*}
\begin{figure*}
  \includegraphics[width=\textwidth]{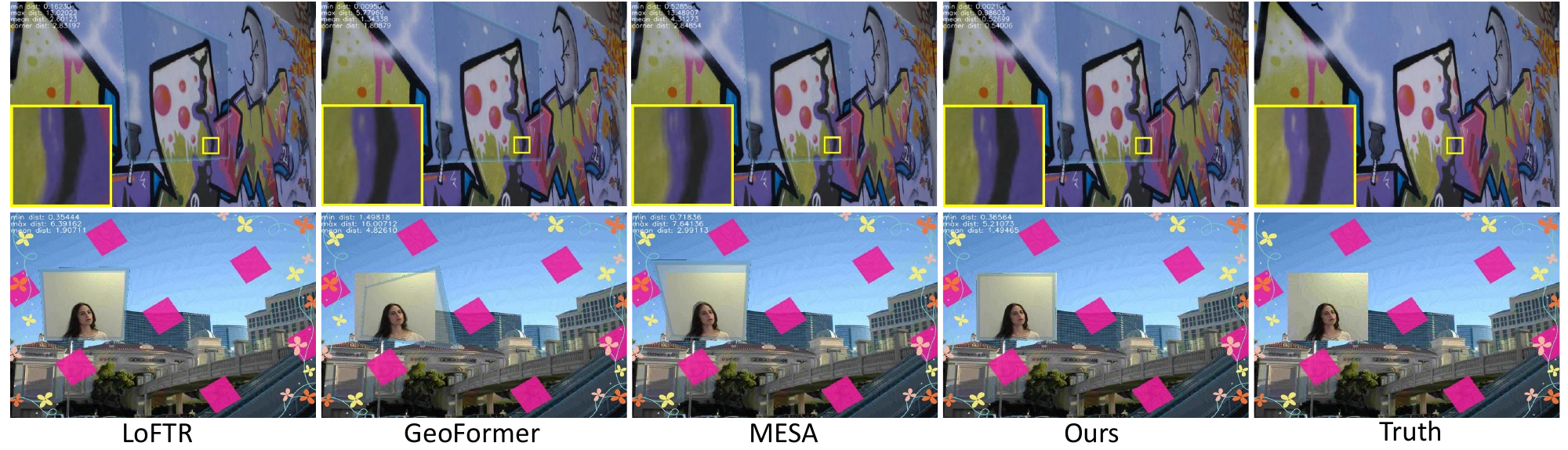}
  \captionsetup{skip=1pt}
  \caption{Qualitative of homography estimation results with LoFTR \cite{sun2021loftr}, GeoFormer \cite{liu2023geometrized}, MESA \cite{zhang2024mesa}, and our SRMatcher. }
  \label{fig_5m}
  \vspace{-4mm}
\end{figure*}

\textbf{Result.} Table \ref{tab:t1m} displays the AUC scores for different methods, SRMatcher consistently exceeds the performance of other baseline methods across all error thresholds. Leveraging the formidable modeling prowess of Transformer, the attention-based matcher demonstrates a notable lead over alternative methodologies. SRMatcher outperforms the previous SOTA GeoFormer in accuracy across various thresholds. SRMatcher demonstrates the most significant enhancement in AUC@1px, scoring 44.3 compared to 49.2. This superiority stems from the introduction of semantic constraints alongside geometric constraints. This integration helps eliminate unsound matches within the same set while strengthening matches based on semantic information. Qualitative results of homography estimation and matching are shown in Figure \ref{fig_4m} and Figure \ref{fig_5m}. From the matching result of ISC-HE, we can see the all the matched point shared same semantic. 

\subsection{Evaluation on Manipulated Images}
The models from Section \ref{section:4.2} are utilized as is, without any further training.

\textbf{Test Data.} We follow \cite{liu2023geometrized} and use the ISC-HE datasets for the evaluation of manipulated images. Images were sourced from the Facebook AI Image Similarity Challenge (ISC) \cite{douze20212021}, where an original image undergoes various edits, like rotations and merges with other images, resulting in highly manipulated images. Given that ISC image pairs are unregistered, they were manually and collectively annotated, yielding 186 registered pairs. Each pair contains at least 8 correspondences. 

\begin{table}[t]
\caption{\textbf{Homography estimation results on ISC-HE.}  The \textbf{best} and \underline{second} results are highlighted.}\label{tab:PAUC_SN}
\centering
\vspace{-3mm}
\resizebox{\linewidth}{!}{
\begin{tabular}{lllllll}
\toprule
\multicolumn{3}{l}{\multirow{3}{*}{Method}} & \multicolumn{3}{c}{Homography est. AUC} \\ \cmidrule(l){4-6} 
\multicolumn{3}{c}{}                        & @3px$\uparrow$       & @5px$\uparrow$        & @10px$\uparrow$        \\ \midrule

\multicolumn{6}{l}{$Detector$-$based$ $matching:$} \\

\multicolumn{3}{l}
{Superpoint~\cite{detone2018superpoint}~\tiny{CVPRW'18}}               & 18.3      & 39.0      & {62.2}     \\
\multicolumn{3}{l}
{R2D2~\cite{revaud2019r2d2}~\tiny{NIPS'19}}               & 18.2      & 39.6      & {62.9}     \\
\multicolumn{3}{l}
{SIFT~\cite{lowe2004distinctive}}               & 19.9      & 42.4      & {65.0}     \\
\multicolumn{3}{l}
{SuperGlue~\cite{sarlin2020superglue}~\tiny{CVPR'20}}               & 19.6      & 42.2      & {66.9}     \\

\midrule
\multicolumn{6}{l}{$Detector$-$free$ $matching:$} \\
\multicolumn{3}{l}{TopicFM~\cite{giang2023topicfm}~\tiny{AAAI'23}}                   & 18.8       & 41.9       & 65.4       \\
\multicolumn{3}{l}{AdaMatcher~\cite{huang2023adaptive}~\tiny{CVPR'23}}                   & 19.0       & 42.6       & 66.5       \\
\multicolumn{3}{l}{CasMTR~\cite{cao2023improving}~\tiny{ICCV'23}}        & {19.4}      & {{43.0}}      & {65.3}       \\ 
\multicolumn{3}{l}{MESA~\cite{zhang2024mesa}~\tiny{CVPR'24}}      & \underline{20.1}      & {{43.6}}      & {68.3}       \\

\midrule
\multicolumn{3}{l}{LoFTR~\cite{sun2021loftr}~\tiny{CVPR'21}}                   & 18.7       & 41.0       & 64.8       \\

\rowcolor[rgb]{.9,.95,.98}\multicolumn{3}{l}{SRMatcher\_LoFTR}        & 19.2$_{+2.67\%}$       & 41.4$_{+0.97\%}$       & 65.1$_{+0.46\%}$       \\ \midrule
\multicolumn{3}{l}{ASpan~\cite{chen2022aspanformer}~\tiny{ECCV'22}}                   & 18.0       & 39.2       & 62.0       \\

\rowcolor[rgb]{.9,.95,.98}\multicolumn{3}{l}{SRMatcher\_ASpan}        & 18.4$_{+2.22\%}$       & 39.5$_{+0.77\%}$       & 62.3$_{+0.48\%}$       \\ \midrule
\multicolumn{3}{l}{GeoFormer~\cite{liu2023geometrized}~\tiny{ICCV'23}}                   & 19.9       & \underline{43.8}       & \underline{68.4}       \\

\rowcolor[rgb]{.9,.95,.98}\multicolumn{3}{l}{SRMatcher\_GeoFormer}        & \textbf{20.5}$_{+3.02\%}$       & \textbf{44.2}$_{+1.01\%}$       & \textbf{68.8}$_{+0.58\%}$       \\ 
\bottomrule
\end{tabular}
}\label{tab:t2m}
\vspace{-5mm}
\end{table}
\textbf{Results.} Table \ref{tab:t2m} illustrates that SRMatcher outperforms all baseline methods, even though the performance margin is smaller compared to the HPatches experiment. This is attributed to the highly challenging nature of ISC-HE. Unlike HPatches, ISC-HE displays forgery characteristics such as watermarks, cutouts, and image stitching, which pose challenges for semantic extraction in accurately capturing the image's semantics.

One thought-provoking finding from Table \ref{tab:t2m} is the detector-based methods notably SuperGlue and SIFT outperform the advanced detector-baseline, i.e. ASpan. This result can be interpreted as follows: ISC-HE images underwent significant modifications, including watermark insertion and background replacement. Unlike detector-free methods that rely on dense feature matching, detector-based approaches are less affected by these alterations. The superior performance of SRMatcher compared to both detector-based and detector-free methods represents the robustness and accuracy of semantic-aware feature matching methods.
\subsection{Evaluation on Relative Pose Estimation}
To demonstrate the applicability of our SRMatcher to other tasks, we test our model on the MegaDepth dataset for relative pose estimation. The models described in Section \ref{section:4.2} are used directly, without undergoing additional training.

\textbf{Test Data.} MegaDepth includes 196 scene reconstructions from 1 million Internet images and 1500 pairs from two distinct scenes chosen for the test set, SRMatcher is tested in 1152 $\times$ 1152 following CasMTR \cite{cao2023improving}.


\textbf{Metrics.} Following \cite{zhang2024mesa}, we present the AUC of the pose error for thresholds (5°, 10°, 20°), where the pose error is determined by the maximal angular error of relative rotation and translation. In our evaluation approach, relative poses are deduced from the essential matrix, which is calculated from feature matching using RANSAC.

\textbf{Results.} As Table \ref{tab:t3m} show, SRMatcher achieves the highest in AUC@5$^\circ$ and AUC@10$^\circ$ shows SRMatcher's strong task adaptability. It's important to note that MESA is designed specifically for relative pose estimation. Furthermore, SRMatcher offers significantly enhancement for GeoFormer, elevating the precision to the state-of-the-art level. This demonstrates that the semantic-aware matching provided by SRMatcher significantly augments the performance of feature matching.
\subsection{Ablation Study}
To assess the efficacy of our design, we perform an extensive ablation study on the components of SRMatcher on HPatches.

\textbf{Semantic-aware fusion block.} In order to demonstrated the significance of incorporating semantic information in the context of homography estimation, as row 1 of Table \ref{tab:t4m} shows, without using semantic information (i.e., w/o SFB) we can see that the performance of AUC@1 decreases from 49.2 to 45.3. This outcome highlights the superior performance of the proposed semantic-aware feature representation learning framework. To show the impact of various fusion techniques utilized in SGIB, we compare our proposed SGIB against spatial attention (i.e., w/o SGIB). Comparing rows 2 and 7, it is evident that the feature produced by our SGIB outperforms the other, achieving the best performance. It indicates that our proposed SGIB by means of executes cross-attention between semantic and image features, fully leveraging the rich semantic information present in VFMs while preserving spatial details. More details are discussed in the supplementary.
 
\begin{table}[t]
\caption{\textbf{Relative pose estimation results $(\%)$ on MegaDepth-1500 benchmark.}  The \textbf{best} and \underline{second} results are highlighted.}\label{tab:PAUC_SN}
\centering
\vspace{-3mm}
\resizebox{\linewidth}{!}{
\begin{tabular}{lllllll}
\toprule
\multicolumn{3}{l}{\multirow{3}{*}{Pose estimation AUC}} & \multicolumn{3}{c}{MegaDepth1500 benchmark} \\ \cmidrule(l){4-6} 
\multicolumn{3}{c}{}                        & AUC@5$^\circ\uparrow$       & AUC@10$^\circ\uparrow$        & AUC@20$^\circ\uparrow$        \\ \midrule
\multicolumn{3}{l}{TopicFM~\cite{giang2023topicfm}~\tiny{AAAI'23}}                   & 32.8       & 47.9       & 63.3       \\
\multicolumn{3}{l}{AdaMatcher~\cite{huang2023adaptive}~\tiny{CVPR'23}}                   & 33.6       & 48.7       & 63.8       \\
\multicolumn{3}{l}{CasMTR~\cite{cao2023improving}~\tiny{ICCV'23}}        & {34.3}      & {{49.9}}      & {64.1}       \\
\multicolumn{3}{l}{MESA~\cite{zhang2024mesa}~\tiny{CVPR'24}}      & \textbf{{36.3}}      & \underline{{51.1}}      & \underline{64.9}       \\

\midrule
\multicolumn{3}{l}{LoFTR~\cite{sun2021loftr}~\tiny{CVPR'21}}                   & 32.3       & 47.8       & 61.4       \\

\multicolumn{3}{l}{SRMatcher\_LoFTR}        & 33.8       & 49.1       & 63.0       \\ \midrule
\multicolumn{3}{l}{ASpan~\cite{chen2022aspanformer}~\tiny{ECCV'22}}                   & 33.6       & 49.2       & 62.8       \\

\multicolumn{3}{l}{SRMatcher\_ASpan}        & 34.9       & 51.0       & 64.4       \\ \midrule
\multicolumn{3}{l}{GeoFormer~\cite{liu2023geometrized}~\tiny{ICCV'23}}                   & 34.2       & {50.5}       & {63.5}       \\

\multicolumn{3}{l}{SRMatcher\_GeoFormer}        & \underline{35.8}       & \textbf{52.0}       & \textbf{65.2}      \\ 
\bottomrule
\end{tabular}
}\label{tab:t3m}
\vspace{-6mm}
\end{table}
\textbf{Semantic Extractors.} To further explore whether the fine-grained semantic features improve the effectiveness of matching results, we choose two separate semantic extractors. As Table \ref{tab:t4m} shows, we compare our approach with the ResNet-50, which is pretrained on ImageNet (i.e., w/ ResNet\_50). The results are reported in row 3. Compared with row 7, it's evident that the semantic features produced by DINOv2 outperform the others in terms of effectiveness. This excellent performance is due to DINOv2's self-supervised training approach, which forces the model to learn image features that remain stable under different transformations, which tend to have a high semantic level. This result highlights the potential of integrating the more refined understanding of semantics, as features enriched with semantics provide robust representational ability.

\textbf{Overlap based fine matching.} To validate the usefulness of our overlap based fine matching, we use fine-matching2 in GeoFormer \cite{liu2023geometrized} as an alternative (i.e., w/o overlap based fine matching). As reported in row 6, we can see that the performance undergoes a significant drop. This result indicates the effects of our overlap based fine matching method.

\textbf{Fusion object.} In our matching network, the semantic features \(S_0\) and \(S_1\) are fused with the features \(\hat{C_0}\) and \(\hat{C_1}\), which are updated through self- and cross-attention. In this experiment, we try to verify the optimal fusion object. An alternative fusion type uses the coarse-level features $C_{0}$ and $C_{1}$. Before feeding into the self-attention and cross-attention mechanisms, we use the SFB to fuse the semantic features with the coarse-level features (i.e., w/ FPN). From row 4 of Table \ref{tab:t4m}, it is noted that fusion with coarse-level features $C_{0}$ and $C_{1}$ results in 47.8 of AUC@1, markedly lower than the 49.2 obtained when fusion with \(\hat{C_0}\) and \(\hat{C_1}\). These results indicate that \(\hat{C_0}\) and \(\hat{C_1}\) updated through the attention mechanism are better suited for fusion due to their enhanced position and context-related local features.

\textbf{Cross-images feature fusion.} To explore the effects of cross-images feature fusion strategy, we only utilize the operation of stage1 within SFB as mentioned in Section \ref{section:3.2} (i.e., w/o cross-images fusion). This means that each image will only be fused with its own semantic features, regardless of semantic information from other images. The results are reported in row 5. Compared with row 7, it clearly show the usefulness of our cross-images feature fusion strategy. It's the same as human intuition, when humans perform matching operations, the area with the same semantics between two images should be paid attention to, because most matching points will be generated in this area. Therefore, it is necessary to pay attention to the area with the same semantics by referring to the semantics of other images.
\captionsetup{
    skip=0.2pt,              
}
\begin{figure}[htbp]
    \vspace{-3.5mm}
    \centering
    \includegraphics[width=0.79\linewidth,trim=10mm 5mm 10mm 20mm, clip]{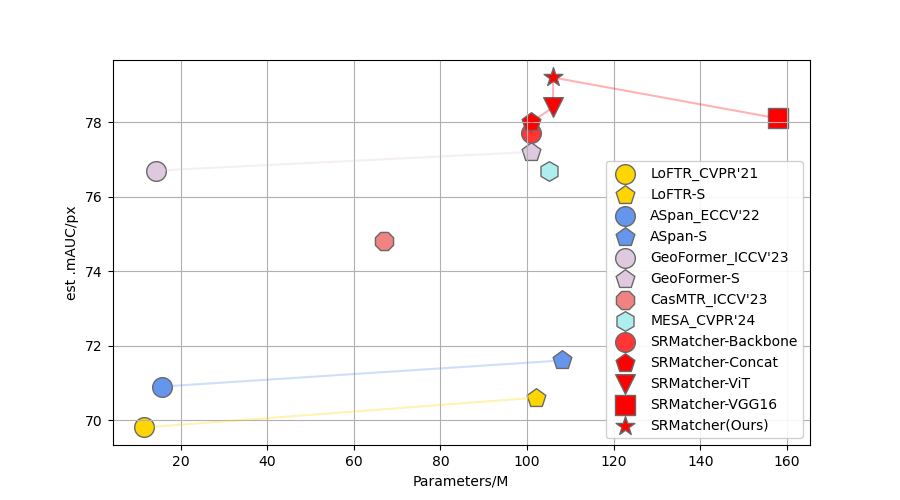}
    \caption{Performance comparison. ``-S'' means use DINOv2 as the backbone to scale the parameters. ``-Backbone'' means use DINOv2 as the backbone to get coarse features. ``-Concat'' means concatenate the DINOv2 and CNN features. ``-ViT'' and ``-VGG16'' mean use different semantic extractors. }
    \label{fig:1}
    \vspace{-6mm}
\end{figure}
\begin{table}[t]

\caption{\textbf{Ablation study.} Various variants of SRMatcher are evaluated for homography estimation on HPatches to highlight the significance of different components.} \label{tab:ASR}
\vspace{-0mm}
\resizebox{\linewidth}{!}{ 
\begin{tabular}{cccccccc}
\toprule
\multicolumn{3}{l}{\multirow{3}{*}{Modification}} & \multicolumn{4}{c}{Homography est. AUC} \\ \cmidrule(l){4-7} 
\multicolumn{3}{c}{}               & @1px$\uparrow$         & @3px$\uparrow$       & @5px$\uparrow$        & @10px$\uparrow$        \\ \midrule
\multicolumn{3}{l}{ w/o~\textit{SFB}}           & 45.3 & 69.3   & 77.8      & 86.0     \\
\multicolumn{3}{l}{ w/o~\textit{SGIB}}          & 47.8 & 70.1   & 78.5      & 86.0     \\
\multicolumn{3}{l}{ w/~\textit{ResNet\_50}} & 47.2     & 69.7       & 78.3          & 86.2        \\
\multicolumn{3}{l}{ w/~\textit{FPN}} & 47.8     & 69.6       & 77.7          & 85.6        \\
\multicolumn{3}{l}{ w/o~\textit{Cross-images fusion}} & 48.1     & 69.8       & 77.5          & 85.9        \\
\multicolumn{3}{l}{ w/o~\textit{Overlap based Fine Matching}} & 47.6     & 70.3       & 78.6          & 86.3        \\

\multicolumn{3}{l}{SRMatcher\_GeoFormer }          & \textbf{49.2} & \textbf{71.2}   &\textbf{79.3}       & \textbf{87.0}     \\
\bottomrule
\end{tabular}
}\label{tab:t4m}
\vspace{-5mm} 
\end{table}

\subsection{Understanding SRMatcher}
SRMatcher focus on \textit{how to integrate semantics into the feature matching model}. Although strong semantic extractor can lead to accuracy improvement. However leveraging the information provided by the semantic extractor is not a trivial task. The evidence to support the above view is that \textbf{1)} Different semantic extractors can be used in our framework to achieve good performance. We conduct the experiments using different semantic extractors and the improvement is universal across all these backbones shown in Figure \ref{fig:1} (i.e. -ViT, -VGG16). \textbf{2)} We compare the naive method to leverage the information of the semantic extractor and our method performs best, which indicates the effectiveness of our method to integrate the information of the semantic extractor, shown in Figure \ref{fig:1} (i.e. -Backbone, -Concat).

Using DINOv2 as a semantic extractor increases the number of parameters in the model, and one possible view is that increasing the number of parameters will naturally improve the model performance. We also conduct extra experiments by controlling the parameters across various methods in Figure \ref{fig:1} (i.e. -S). Our proposed SRMatcher achieves the best performance among the scaled version methods and the latest work with similar parameters.

\section{Limitation and Discussion}
Firstly, the inclusion of an additional semantic extractor and Seman-tic-aware Fusion Block (SFB) increases the computational complexity of SRMatcher compared to previous methods. Consequently, when processing large input images, the inference time of the model is prolonged. However, computational efficiency can be enhanced by optimizing the model’s structure.

The generalization ability of the model is another critical aspect. When confronted with scientific data sets, such as medical images, DINOv2 may be unable to accurately identify image semantics. This is due to the limited training data set of DINOv2, which prevents generalization to more scenarios. In the future, constructing more extensive datasets for training could significantly improve its semantic extraction capabilities.

\section{CONCLUSION}
This work has proposed a novel detector-free feature matching method for homography estimation, named SRMatcher. 
We have discovered that previous matching networks only utilized coarse-grained semantic features, which prevents the full utilization of the knowledge that semantic extraction networks can offer and lacks adaptability with other tasks. Different from previous works, we explore a new way to integrate semantic information with the matching network. Specifically, we incorporate fine-grained image semantics derived from vision foundation models (VFMs), using our proposed Semantic-aware Fusion Block (SFB) to conduct cross-images feature interaction. This encourages our SRMatcher to to learn integrated semantic feature representation acquiring more accurate and realistic outcomes. Additionally, our SRMatcher can be seamlessly integrated into the majority of LoFTR-based feature matching methods, where it consistently delivers outstanding results. Through comprehensive experimentation, we have explored the impact of our discoveries and highlighted the advantages of our proposed SRMatcher. We are confident that SRMatcher will provide new insights to the homography estimation community.

\section{Acknowledgements}
This work was supported in part by National Science and Technology Major Project under Grant 2023ZD0121300, National Natural Science Foundation of China under Grants 62088102, 12326608 and 62106192, Natural Science Foundation of Shaanxi Province under Grant 2022JC-41, and Fundamental Research Funds for the Central Universities under Grant XTR042021005.

\bibliographystyle{ACM-Reference-Format}
\bibliography{sample-base}


\begin{thebibliography}{53}


\ifx \showCODEN    \undefined \def \showCODEN     #1{\unskip}     \fi
\ifx \showDOI      \undefined \def \showDOI       #1{#1}\fi
\ifx \showISBNx    \undefined \def \showISBNx     #1{\unskip}     \fi
\ifx \showISBNxiii \undefined \def \showISBNxiii  #1{\unskip}     \fi
\ifx \showISSN     \undefined \def \showISSN      #1{\unskip}     \fi
\ifx \showLCCN     \undefined \def \showLCCN      #1{\unskip}     \fi
\ifx \shownote     \undefined \def \shownote      #1{#1}          \fi
\ifx \showarticletitle \undefined \def \showarticletitle #1{#1}   \fi
\ifx \showURL      \undefined \def \showURL       {\relax}        \fi
\providecommand\bibfield[2]{#2}
\providecommand\bibinfo[2]{#2}
\providecommand\natexlab[1]{#1}
\providecommand\showeprint[2][]{arXiv:#2}

\bibitem[Arandjelovi{\'c} and Zisserman(2012)]%
        {arandjelovic2012three}
\bibfield{author}{\bibinfo{person}{Relja Arandjelovi{\'c}} {and} \bibinfo{person}{Andrew Zisserman}.} \bibinfo{year}{2012}\natexlab{}.
\newblock \showarticletitle{Three things everyone should know to improve object retrieval}. In \bibinfo{booktitle}{\emph{2012 IEEE conference on computer vision and pattern recognition}}. IEEE, \bibinfo{pages}{2911--2918}.
\newblock


\bibitem[Balntas et~al\mbox{.}(2017)]%
        {balntas2017hpatches}
\bibfield{author}{\bibinfo{person}{Vassileios Balntas}, \bibinfo{person}{Karel Lenc}, \bibinfo{person}{Andrea Vedaldi}, {and} \bibinfo{person}{Krystian Mikolajczyk}.} \bibinfo{year}{2017}\natexlab{}.
\newblock \showarticletitle{HPatches: A benchmark and evaluation of handcrafted and learned local descriptors}. In \bibinfo{booktitle}{\emph{Proceedings of the IEEE conference on computer vision and pattern recognition}}. \bibinfo{pages}{5173--5182}.
\newblock


\bibitem[Cao and Fu(2023)]%
        {cao2023improving}
\bibfield{author}{\bibinfo{person}{Chenjie Cao} {and} \bibinfo{person}{Yanwei Fu}.} \bibinfo{year}{2023}\natexlab{}.
\newblock \showarticletitle{Improving transformer-based image matching by cascaded capturing spatially informative keypoints}. In \bibinfo{booktitle}{\emph{Proceedings of the IEEE/CVF International Conference on Computer Vision}}. \bibinfo{pages}{12129--12139}.
\newblock


\bibitem[Chang et~al\mbox{.}(2023)]%
        {chang2023structured}
\bibfield{author}{\bibinfo{person}{Jiahao Chang}, \bibinfo{person}{Jiahuan Yu}, {and} \bibinfo{person}{Tianzhu Zhang}.} \bibinfo{year}{2023}\natexlab{}.
\newblock \showarticletitle{Structured Epipolar Matcher for Local Feature Matching}. In \bibinfo{booktitle}{\emph{Proceedings of the IEEE/CVF Conference on Computer Vision and Pattern Recognition}}. \bibinfo{pages}{6176--6185}.
\newblock


\bibitem[Chen et~al\mbox{.}(2022)]%
        {chen2022aspanformer}
\bibfield{author}{\bibinfo{person}{Hongkai Chen}, \bibinfo{person}{Zixin Luo}, \bibinfo{person}{Lei Zhou}, \bibinfo{person}{Yurun Tian}, \bibinfo{person}{Mingmin Zhen}, \bibinfo{person}{Tian Fang}, \bibinfo{person}{David Mckinnon}, \bibinfo{person}{Yanghai Tsin}, {and} \bibinfo{person}{Long Quan}.} \bibinfo{year}{2022}\natexlab{}.
\newblock \showarticletitle{Aspanformer: Detector-free image matching with adaptive span transformer}. In \bibinfo{booktitle}{\emph{European Conference on Computer Vision}}. Springer, \bibinfo{pages}{20--36}.
\newblock


\bibitem[DeTone et~al\mbox{.}(2018)]%
        {detone2018superpoint}
\bibfield{author}{\bibinfo{person}{Daniel DeTone}, \bibinfo{person}{Tomasz Malisiewicz}, {and} \bibinfo{person}{Andrew Rabinovich}.} \bibinfo{year}{2018}\natexlab{}.
\newblock \showarticletitle{Superpoint: Self-supervised interest point detection and description}. In \bibinfo{booktitle}{\emph{Proceedings of the IEEE conference on computer vision and pattern recognition workshops}}. \bibinfo{pages}{224--236}.
\newblock


\bibitem[Devlin et~al\mbox{.}(2018)]%
        {devlin2018bert}
\bibfield{author}{\bibinfo{person}{Jacob Devlin}, \bibinfo{person}{Ming-Wei Chang}, \bibinfo{person}{Kenton Lee}, {and} \bibinfo{person}{Kristina Toutanova}.} \bibinfo{year}{2018}\natexlab{}.
\newblock \showarticletitle{Bert: Pre-training of deep bidirectional transformers for language understanding}.
\newblock \bibinfo{journal}{\emph{arXiv preprint arXiv:1810.04805}} (\bibinfo{year}{2018}).
\newblock


\bibitem[Dosovitskiy et~al\mbox{.}(2020)]%
        {dosovitskiy2020image}
\bibfield{author}{\bibinfo{person}{Alexey Dosovitskiy}, \bibinfo{person}{Lucas Beyer}, \bibinfo{person}{Alexander Kolesnikov}, \bibinfo{person}{Dirk Weissenborn}, \bibinfo{person}{Xiaohua Zhai}, \bibinfo{person}{Thomas Unterthiner}, \bibinfo{person}{Mostafa Dehghani}, \bibinfo{person}{Matthias Minderer}, \bibinfo{person}{Georg Heigold}, \bibinfo{person}{Sylvain Gelly}, {et~al\mbox{.}}} \bibinfo{year}{2020}\natexlab{}.
\newblock \showarticletitle{An image is worth 16x16 words: Transformers for image recognition at scale}.
\newblock \bibinfo{journal}{\emph{arXiv preprint arXiv:2010.11929}} (\bibinfo{year}{2020}).
\newblock


\bibitem[Douze et~al\mbox{.}(2021)]%
        {douze20212021}
\bibfield{author}{\bibinfo{person}{Matthijs Douze}, \bibinfo{person}{Giorgos Tolias}, \bibinfo{person}{Ed Pizzi}, \bibinfo{person}{Zo{\"e} Papakipos}, \bibinfo{person}{Lowik Chanussot}, \bibinfo{person}{Filip Radenovic}, \bibinfo{person}{Tomas Jenicek}, \bibinfo{person}{Maxim Maximov}, \bibinfo{person}{Laura Leal-Taix{\'e}}, \bibinfo{person}{Ismail Elezi}, {et~al\mbox{.}}} \bibinfo{year}{2021}\natexlab{}.
\newblock \showarticletitle{The 2021 image similarity dataset and challenge}.
\newblock \bibinfo{journal}{\emph{arXiv preprint arXiv:2106.09672}} (\bibinfo{year}{2021}).
\newblock


\bibitem[Edstedt et~al\mbox{.}(2023a)]%
        {edstedt2023dkm}
\bibfield{author}{\bibinfo{person}{Johan Edstedt}, \bibinfo{person}{Ioannis Athanasiadis}, \bibinfo{person}{M{\aa}rten Wadenb{\"a}ck}, {and} \bibinfo{person}{Michael Felsberg}.} \bibinfo{year}{2023}\natexlab{a}.
\newblock \showarticletitle{DKM: Dense kernelized feature matching for geometry estimation}. In \bibinfo{booktitle}{\emph{Proceedings of the IEEE/CVF Conference on Computer Vision and Pattern Recognition}}. \bibinfo{pages}{17765--17775}.
\newblock


\bibitem[Edstedt et~al\mbox{.}(2023b)]%
        {edstedt2023roma}
\bibfield{author}{\bibinfo{person}{Johan Edstedt}, \bibinfo{person}{Qiyu Sun}, \bibinfo{person}{Georg B{\"o}kman}, \bibinfo{person}{M{\aa}rten Wadenb{\"a}ck}, {and} \bibinfo{person}{Michael Felsberg}.} \bibinfo{year}{2023}\natexlab{b}.
\newblock \showarticletitle{RoMa: Revisiting Robust Losses for Dense Feature Matching}.
\newblock \bibinfo{journal}{\emph{arXiv preprint arXiv:2305.15404}} (\bibinfo{year}{2023}).
\newblock


\bibitem[Gelfand et~al\mbox{.}(2010)]%
        {gelfand2010multi}
\bibfield{author}{\bibinfo{person}{Natasha Gelfand}, \bibinfo{person}{Andrew Adams}, \bibinfo{person}{Sung~Hee Park}, {and} \bibinfo{person}{Kari Pulli}.} \bibinfo{year}{2010}\natexlab{}.
\newblock \showarticletitle{Multi-exposure imaging on mobile devices}. In \bibinfo{booktitle}{\emph{Proceedings of the 18th ACM international conference on Multimedia}}. \bibinfo{pages}{823--826}.
\newblock


\bibitem[Giang et~al\mbox{.}(2023)]%
        {giang2023topicfm}
\bibfield{author}{\bibinfo{person}{Khang~Truong Giang}, \bibinfo{person}{Soohwan Song}, {and} \bibinfo{person}{Sungho Jo}.} \bibinfo{year}{2023}\natexlab{}.
\newblock \showarticletitle{TopicFM: Robust and interpretable topic-assisted feature matching}. In \bibinfo{booktitle}{\emph{Proceedings of the AAAI conference on artificial intelligence}}, Vol.~\bibinfo{volume}{37}. \bibinfo{pages}{2447--2455}.
\newblock


\bibitem[Guo et~al\mbox{.}(2016)]%
        {guo2016joint}
\bibfield{author}{\bibinfo{person}{Heng Guo}, \bibinfo{person}{Shuaicheng Liu}, \bibinfo{person}{Tong He}, \bibinfo{person}{Shuyuan Zhu}, \bibinfo{person}{Bing Zeng}, {and} \bibinfo{person}{Moncef Gabbouj}.} \bibinfo{year}{2016}\natexlab{}.
\newblock \showarticletitle{Joint video stitching and stabilization from moving cameras}.
\newblock \bibinfo{journal}{\emph{IEEE Transactions on Image Processing}} \bibinfo{volume}{25}, \bibinfo{number}{11} (\bibinfo{year}{2016}), \bibinfo{pages}{5491--5503}.
\newblock


\bibitem[He et~al\mbox{.}(2022)]%
        {he2022masked}
\bibfield{author}{\bibinfo{person}{Kaiming He}, \bibinfo{person}{Xinlei Chen}, \bibinfo{person}{Saining Xie}, \bibinfo{person}{Yanghao Li}, \bibinfo{person}{Piotr Doll{\'a}r}, {and} \bibinfo{person}{Ross Girshick}.} \bibinfo{year}{2022}\natexlab{}.
\newblock \showarticletitle{Masked autoencoders are scalable vision learners}. In \bibinfo{booktitle}{\emph{Proceedings of the IEEE/CVF conference on computer vision and pattern recognition}}. \bibinfo{pages}{16000--16009}.
\newblock


\bibitem[He et~al\mbox{.}(2016)]%
        {he2016deep}
\bibfield{author}{\bibinfo{person}{Kaiming He}, \bibinfo{person}{Xiangyu Zhang}, \bibinfo{person}{Shaoqing Ren}, {and} \bibinfo{person}{Jian Sun}.} \bibinfo{year}{2016}\natexlab{}.
\newblock \showarticletitle{Deep residual learning for image recognition}. In \bibinfo{booktitle}{\emph{Proceedings of the IEEE conference on computer vision and pattern recognition}}. \bibinfo{pages}{770--778}.
\newblock


\bibitem[Hong et~al\mbox{.}(2022)]%
        {hong2022unsupervised}
\bibfield{author}{\bibinfo{person}{Mingbo Hong}, \bibinfo{person}{Yuhang Lu}, \bibinfo{person}{Nianjin Ye}, \bibinfo{person}{Chunyu Lin}, \bibinfo{person}{Qijun Zhao}, {and} \bibinfo{person}{Shuaicheng Liu}.} \bibinfo{year}{2022}\natexlab{}.
\newblock \showarticletitle{Unsupervised homography estimation with coplanarity-aware gan}. In \bibinfo{booktitle}{\emph{Proceedings of the IEEE/CVF conference on computer vision and pattern recognition}}. \bibinfo{pages}{17663--17672}.
\newblock


\bibitem[Huang et~al\mbox{.}(2023)]%
        {huang2023adaptive}
\bibfield{author}{\bibinfo{person}{Dihe Huang}, \bibinfo{person}{Ying Chen}, \bibinfo{person}{Yong Liu}, \bibinfo{person}{Jianlin Liu}, \bibinfo{person}{Shang Xu}, \bibinfo{person}{Wenlong Wu}, \bibinfo{person}{Yikang Ding}, \bibinfo{person}{Fan Tang}, {and} \bibinfo{person}{Chengjie Wang}.} \bibinfo{year}{2023}\natexlab{}.
\newblock \showarticletitle{Adaptive assignment for geometry aware local feature matching}. In \bibinfo{booktitle}{\emph{Proceedings of the IEEE/CVF Conference on Computer Vision and Pattern Recognition}}. \bibinfo{pages}{5425--5434}.
\newblock


\bibitem[Jiang et~al\mbox{.}(2021)]%
        {jiang2021cotr}
\bibfield{author}{\bibinfo{person}{Wei Jiang}, \bibinfo{person}{Eduard Trulls}, \bibinfo{person}{Jan Hosang}, \bibinfo{person}{Andrea Tagliasacchi}, {and} \bibinfo{person}{Kwang~Moo Yi}.} \bibinfo{year}{2021}\natexlab{}.
\newblock \showarticletitle{Cotr: Correspondence transformer for matching across images}. In \bibinfo{booktitle}{\emph{Proceedings of the IEEE/CVF International Conference on Computer Vision}}. \bibinfo{pages}{6207--6217}.
\newblock


\bibitem[Katharopoulos et~al\mbox{.}(2020)]%
        {katharopoulos2020transformers}
\bibfield{author}{\bibinfo{person}{Angelos Katharopoulos}, \bibinfo{person}{Apoorv Vyas}, \bibinfo{person}{Nikolaos Pappas}, {and} \bibinfo{person}{Fran{\c{c}}ois Fleuret}.} \bibinfo{year}{2020}\natexlab{}.
\newblock \showarticletitle{Transformers are rnns: Fast autoregressive transformers with linear attention}. In \bibinfo{booktitle}{\emph{International conference on machine learning}}. PMLR, \bibinfo{pages}{5156--5165}.
\newblock


\bibitem[Kirillov et~al\mbox{.}(2023)]%
        {kirillov2023segment}
\bibfield{author}{\bibinfo{person}{Alexander Kirillov}, \bibinfo{person}{Eric Mintun}, \bibinfo{person}{Nikhila Ravi}, \bibinfo{person}{Hanzi Mao}, \bibinfo{person}{Chloe Rolland}, \bibinfo{person}{Laura Gustafson}, \bibinfo{person}{Tete Xiao}, \bibinfo{person}{Spencer Whitehead}, \bibinfo{person}{Alexander~C Berg}, \bibinfo{person}{Wan-Yen Lo}, {et~al\mbox{.}}} \bibinfo{year}{2023}\natexlab{}.
\newblock \showarticletitle{Segment anything}. In \bibinfo{booktitle}{\emph{Proceedings of the IEEE/CVF International Conference on Computer Vision}}. \bibinfo{pages}{4015--4026}.
\newblock


\bibitem[Li et~al\mbox{.}(2024)]%
        {li2024sed}
\bibfield{author}{\bibinfo{person}{Bingchen Li}, \bibinfo{person}{Xin Li}, \bibinfo{person}{Hanxin Zhu}, \bibinfo{person}{Yeying Jin}, \bibinfo{person}{Ruoyu Feng}, \bibinfo{person}{Zhizheng Zhang}, {and} \bibinfo{person}{Zhibo Chen}.} \bibinfo{year}{2024}\natexlab{}.
\newblock \showarticletitle{SeD: Semantic-Aware Discriminator for Image Super-Resolution}.
\newblock \bibinfo{journal}{\emph{arXiv preprint arXiv:2402.19387}} (\bibinfo{year}{2024}).
\newblock


\bibitem[Li and Snavely(2018)]%
        {li2018megadepth}
\bibfield{author}{\bibinfo{person}{Zhengqi Li} {and} \bibinfo{person}{Noah Snavely}.} \bibinfo{year}{2018}\natexlab{}.
\newblock \showarticletitle{Megadepth: Learning single-view depth prediction from internet photos}. In \bibinfo{booktitle}{\emph{Proceedings of the IEEE conference on computer vision and pattern recognition}}. \bibinfo{pages}{2041--2050}.
\newblock


\bibitem[Lin et~al\mbox{.}(2017)]%
        {lin2017feature}
\bibfield{author}{\bibinfo{person}{Tsung-Yi Lin}, \bibinfo{person}{Piotr Doll{\'a}r}, \bibinfo{person}{Ross Girshick}, \bibinfo{person}{Kaiming He}, \bibinfo{person}{Bharath Hariharan}, {and} \bibinfo{person}{Serge Belongie}.} \bibinfo{year}{2017}\natexlab{}.
\newblock \showarticletitle{Feature pyramid networks for object detection}. In \bibinfo{booktitle}{\emph{Proceedings of the IEEE conference on computer vision and pattern recognition}}. \bibinfo{pages}{2117--2125}.
\newblock


\bibitem[Liu et~al\mbox{.}(2024)]%
        {liu2024visual}
\bibfield{author}{\bibinfo{person}{Haotian Liu}, \bibinfo{person}{Chunyuan Li}, \bibinfo{person}{Qingyang Wu}, {and} \bibinfo{person}{Yong~Jae Lee}.} \bibinfo{year}{2024}\natexlab{}.
\newblock \showarticletitle{Visual instruction tuning}.
\newblock \bibinfo{journal}{\emph{Advances in neural information processing systems}}  \bibinfo{volume}{36} (\bibinfo{year}{2024}).
\newblock


\bibitem[Liu and Li(2023)]%
        {liu2023geometrized}
\bibfield{author}{\bibinfo{person}{Jiazhen Liu} {and} \bibinfo{person}{Xirong Li}.} \bibinfo{year}{2023}\natexlab{}.
\newblock \showarticletitle{Geometrized Transformer for Self-Supervised Homography Estimation}. In \bibinfo{booktitle}{\emph{Proceedings of the IEEE/CVF International Conference on Computer Vision}}. \bibinfo{pages}{9556--9565}.
\newblock


\bibitem[Liu et~al\mbox{.}(2023)]%
        {liu2023matcher}
\bibfield{author}{\bibinfo{person}{Yang Liu}, \bibinfo{person}{Muzhi Zhu}, \bibinfo{person}{Hengtao Li}, \bibinfo{person}{Hao Chen}, \bibinfo{person}{Xinlong Wang}, {and} \bibinfo{person}{Chunhua Shen}.} \bibinfo{year}{2023}\natexlab{}.
\newblock \showarticletitle{Matcher: Segment anything with one shot using all-purpose feature matching}.
\newblock \bibinfo{journal}{\emph{arXiv preprint arXiv:2305.13310}} (\bibinfo{year}{2023}).
\newblock


\bibitem[Lowe(2004)]%
        {lowe2004distinctive}
\bibfield{author}{\bibinfo{person}{David~G Lowe}.} \bibinfo{year}{2004}\natexlab{}.
\newblock \showarticletitle{Distinctive image features from scale-invariant keypoints}.
\newblock \bibinfo{journal}{\emph{International journal of computer vision}}  \bibinfo{volume}{60} (\bibinfo{year}{2004}), \bibinfo{pages}{91--110}.
\newblock


\bibitem[Mur-Artal et~al\mbox{.}(2015)]%
        {mur2015orb}
\bibfield{author}{\bibinfo{person}{Raul Mur-Artal}, \bibinfo{person}{Jose Maria~Martinez Montiel}, {and} \bibinfo{person}{Juan~D Tardos}.} \bibinfo{year}{2015}\natexlab{}.
\newblock \showarticletitle{ORB-SLAM: a versatile and accurate monocular SLAM system}.
\newblock \bibinfo{journal}{\emph{IEEE transactions on robotics}} \bibinfo{volume}{31}, \bibinfo{number}{5} (\bibinfo{year}{2015}), \bibinfo{pages}{1147--1163}.
\newblock


\bibitem[Oquab et~al\mbox{.}(2023)]%
        {oquab2023dinov2}
\bibfield{author}{\bibinfo{person}{Maxime Oquab}, \bibinfo{person}{Timoth{\'e}e Darcet}, \bibinfo{person}{Th{\'e}o Moutakanni}, \bibinfo{person}{Huy Vo}, \bibinfo{person}{Marc Szafraniec}, \bibinfo{person}{Vasil Khalidov}, \bibinfo{person}{Pierre Fernandez}, \bibinfo{person}{Daniel Haziza}, \bibinfo{person}{Francisco Massa}, \bibinfo{person}{Alaaeldin El-Nouby}, {et~al\mbox{.}}} \bibinfo{year}{2023}\natexlab{}.
\newblock \showarticletitle{Dinov2: Learning robust visual features without supervision}.
\newblock \bibinfo{journal}{\emph{arXiv preprint arXiv:2304.07193}} (\bibinfo{year}{2023}).
\newblock


\bibitem[Philbin et~al\mbox{.}(2007)]%
        {philbin2007object}
\bibfield{author}{\bibinfo{person}{James Philbin}, \bibinfo{person}{Ondrej Chum}, \bibinfo{person}{Michael Isard}, \bibinfo{person}{Josef Sivic}, {and} \bibinfo{person}{Andrew Zisserman}.} \bibinfo{year}{2007}\natexlab{}.
\newblock \showarticletitle{Object retrieval with large vocabularies and fast spatial matching}. In \bibinfo{booktitle}{\emph{2007 IEEE conference on computer vision and pattern recognition}}. IEEE, \bibinfo{pages}{1--8}.
\newblock


\bibitem[Philbin et~al\mbox{.}(2008)]%
        {philbin2008lost}
\bibfield{author}{\bibinfo{person}{James Philbin}, \bibinfo{person}{Ondrej Chum}, \bibinfo{person}{Michael Isard}, \bibinfo{person}{Josef Sivic}, {and} \bibinfo{person}{Andrew Zisserman}.} \bibinfo{year}{2008}\natexlab{}.
\newblock \showarticletitle{Lost in quantization: Improving particular object retrieval in large scale image databases}. In \bibinfo{booktitle}{\emph{2008 IEEE conference on computer vision and pattern recognition}}. IEEE, \bibinfo{pages}{1--8}.
\newblock


\bibitem[Radford et~al\mbox{.}(2021)]%
        {radford2021learning}
\bibfield{author}{\bibinfo{person}{Alec Radford}, \bibinfo{person}{Jong~Wook Kim}, \bibinfo{person}{Chris Hallacy}, \bibinfo{person}{Aditya Ramesh}, \bibinfo{person}{Gabriel Goh}, \bibinfo{person}{Sandhini Agarwal}, \bibinfo{person}{Girish Sastry}, \bibinfo{person}{Amanda Askell}, \bibinfo{person}{Pamela Mishkin}, \bibinfo{person}{Jack Clark}, {et~al\mbox{.}}} \bibinfo{year}{2021}\natexlab{}.
\newblock \showarticletitle{Learning transferable visual models from natural language supervision}. In \bibinfo{booktitle}{\emph{International conference on machine learning}}. PMLR, \bibinfo{pages}{8748--8763}.
\newblock


\bibitem[Revaud et~al\mbox{.}(2019)]%
        {revaud2019r2d2}
\bibfield{author}{\bibinfo{person}{Jerome Revaud}, \bibinfo{person}{Cesar De~Souza}, \bibinfo{person}{Martin Humenberger}, {and} \bibinfo{person}{Philippe Weinzaepfel}.} \bibinfo{year}{2019}\natexlab{}.
\newblock \showarticletitle{R2d2: Reliable and repeatable detector and descriptor}.
\newblock \bibinfo{journal}{\emph{Advances in neural information processing systems}}  \bibinfo{volume}{32} (\bibinfo{year}{2019}).
\newblock


\bibitem[Rublee et~al\mbox{.}(2011)]%
        {rublee2011orb}
\bibfield{author}{\bibinfo{person}{Ethan Rublee}, \bibinfo{person}{Vincent Rabaud}, \bibinfo{person}{Kurt Konolige}, {and} \bibinfo{person}{Gary Bradski}.} \bibinfo{year}{2011}\natexlab{}.
\newblock \showarticletitle{ORB: An efficient alternative to SIFT or SURF}. In \bibinfo{booktitle}{\emph{2011 International conference on computer vision}}. Ieee, \bibinfo{pages}{2564--2571}.
\newblock


\bibitem[Sarlin et~al\mbox{.}(2020)]%
        {sarlin2020superglue}
\bibfield{author}{\bibinfo{person}{Paul-Edouard Sarlin}, \bibinfo{person}{Daniel DeTone}, \bibinfo{person}{Tomasz Malisiewicz}, {and} \bibinfo{person}{Andrew Rabinovich}.} \bibinfo{year}{2020}\natexlab{}.
\newblock \showarticletitle{Superglue: Learning feature matching with graph neural networks}. In \bibinfo{booktitle}{\emph{Proceedings of the IEEE/CVF conference on computer vision and pattern recognition}}. \bibinfo{pages}{4938--4947}.
\newblock


\bibitem[Sun et~al\mbox{.}(2021)]%
        {sun2021loftr}
\bibfield{author}{\bibinfo{person}{Jiaming Sun}, \bibinfo{person}{Zehong Shen}, \bibinfo{person}{Yuang Wang}, \bibinfo{person}{Hujun Bao}, {and} \bibinfo{person}{Xiaowei Zhou}.} \bibinfo{year}{2021}\natexlab{}.
\newblock \showarticletitle{LoFTR: Detector-free local feature matching with transformers}. In \bibinfo{booktitle}{\emph{Proceedings of the IEEE/CVF conference on computer vision and pattern recognition}}. \bibinfo{pages}{8922--8931}.
\newblock


\bibitem[Tang et~al\mbox{.}(2022)]%
        {tang2022quadtree}
\bibfield{author}{\bibinfo{person}{Shitao Tang}, \bibinfo{person}{Jiahui Zhang}, \bibinfo{person}{Siyu Zhu}, {and} \bibinfo{person}{Ping Tan}.} \bibinfo{year}{2022}\natexlab{}.
\newblock \showarticletitle{Quadtree attention for vision transformers}.
\newblock \bibinfo{journal}{\emph{arXiv preprint arXiv:2201.02767}} (\bibinfo{year}{2022}).
\newblock


\bibitem[Truong~Giang et~al\mbox{.}(2023)]%
        {truong2023topicfm+}
\bibfield{author}{\bibinfo{person}{Khang Truong~Giang}, \bibinfo{person}{Soohwan Song}, {and} \bibinfo{person}{Sungho Jo}.} \bibinfo{year}{2023}\natexlab{}.
\newblock \showarticletitle{TopicFM+: Boosting Accuracy and Efficiency of Topic-Assisted Feature Matching}.
\newblock \bibinfo{journal}{\emph{arXiv e-prints}} (\bibinfo{year}{2023}), \bibinfo{pages}{arXiv--2307}.
\newblock


\bibitem[Vaswani et~al\mbox{.}(2017)]%
        {vaswani2017attention}
\bibfield{author}{\bibinfo{person}{Ashish Vaswani}, \bibinfo{person}{Noam Shazeer}, \bibinfo{person}{Niki Parmar}, \bibinfo{person}{Jakob Uszkoreit}, \bibinfo{person}{Llion Jones}, \bibinfo{person}{Aidan~N Gomez}, \bibinfo{person}{{\L}ukasz Kaiser}, {and} \bibinfo{person}{Illia Polosukhin}.} \bibinfo{year}{2017}\natexlab{}.
\newblock \showarticletitle{Attention is all you need}.
\newblock \bibinfo{journal}{\emph{Advances in neural information processing systems}}  \bibinfo{volume}{30} (\bibinfo{year}{2017}).
\newblock


\bibitem[Wolfe et~al\mbox{.}(2011)]%
        {wolfe2011visual}
\bibfield{author}{\bibinfo{person}{Jeremy~M Wolfe}, \bibinfo{person}{Melissa L-H V{\~o}}, \bibinfo{person}{Karla~K Evans}, {and} \bibinfo{person}{Michelle~R Greene}.} \bibinfo{year}{2011}\natexlab{}.
\newblock \showarticletitle{Visual search in scenes involves selective and nonselective pathways}.
\newblock \bibinfo{journal}{\emph{Trends in cognitive sciences}} \bibinfo{volume}{15}, \bibinfo{number}{2} (\bibinfo{year}{2011}), \bibinfo{pages}{77--84}.
\newblock


\bibitem[Wu et~al\mbox{.}(2014)]%
        {wu2014guidance}
\bibfield{author}{\bibinfo{person}{Chia-Chien Wu}, \bibinfo{person}{Farahnaz~Ahmed Wick}, {and} \bibinfo{person}{Marc Pomplun}.} \bibinfo{year}{2014}\natexlab{}.
\newblock \showarticletitle{Guidance of visual attention by semantic information in real-world scenes}.
\newblock \bibinfo{journal}{\emph{Frontiers in psychology}}  \bibinfo{volume}{5} (\bibinfo{year}{2014}), \bibinfo{pages}{54}.
\newblock


\bibitem[Wu et~al\mbox{.}(2023b)]%
        {wu2023seesr}
\bibfield{author}{\bibinfo{person}{Rongyuan Wu}, \bibinfo{person}{Tao Yang}, \bibinfo{person}{Lingchen Sun}, \bibinfo{person}{Zhengqiang Zhang}, \bibinfo{person}{Shuai Li}, {and} \bibinfo{person}{Lei Zhang}.} \bibinfo{year}{2023}\natexlab{b}.
\newblock \showarticletitle{SeeSR: Towards Semantics-Aware Real-World Image Super-Resolution}.
\newblock \bibinfo{journal}{\emph{arXiv preprint arXiv:2311.16518}} (\bibinfo{year}{2023}).
\newblock


\bibitem[Wu et~al\mbox{.}(2023a)]%
        {wu2023learning}
\bibfield{author}{\bibinfo{person}{Yuhui Wu}, \bibinfo{person}{Chen Pan}, \bibinfo{person}{Guoqing Wang}, \bibinfo{person}{Yang Yang}, \bibinfo{person}{Jiwei Wei}, \bibinfo{person}{Chongyi Li}, {and} \bibinfo{person}{Heng~Tao Shen}.} \bibinfo{year}{2023}\natexlab{a}.
\newblock \showarticletitle{Learning semantic-aware knowledge guidance for low-light image enhancement}. In \bibinfo{booktitle}{\emph{Proceedings of the IEEE/CVF Conference on Computer Vision and Pattern Recognition}}. \bibinfo{pages}{1662--1671}.
\newblock


\bibitem[Yang et~al\mbox{.}(2024)]%
        {yang2024depth}
\bibfield{author}{\bibinfo{person}{Lihe Yang}, \bibinfo{person}{Bingyi Kang}, \bibinfo{person}{Zilong Huang}, \bibinfo{person}{Xiaogang Xu}, \bibinfo{person}{Jiashi Feng}, {and} \bibinfo{person}{Hengshuang Zhao}.} \bibinfo{year}{2024}\natexlab{}.
\newblock \showarticletitle{Depth anything: Unleashing the power of large-scale unlabeled data}.
\newblock \bibinfo{journal}{\emph{arXiv preprint arXiv:2401.10891}} (\bibinfo{year}{2024}).
\newblock


\bibitem[Yu et~al\mbox{.}(2023)]%
        {yu2023astr}
\bibfield{author}{\bibinfo{person}{Jiahuan Yu}, \bibinfo{person}{Jiahao Chang}, \bibinfo{person}{Jianfeng He}, \bibinfo{person}{Tianzhu Zhang}, \bibinfo{person}{Jiyang Yu}, {and} \bibinfo{person}{Wu Feng}.} \bibinfo{year}{2023}\natexlab{}.
\newblock \showarticletitle{ASTR: Adaptive spot-guided transformer for consistent local feature matching}. In \bibinfo{booktitle}{\emph{The IEEE/CVF Computer Vision and Pattern Recognition Conference (CVPR)}}, Vol.~\bibinfo{volume}{7}.
\newblock


\bibitem[Zaragoza et~al\mbox{.}(2013)]%
        {zaragoza2013projective}
\bibfield{author}{\bibinfo{person}{Julio Zaragoza}, \bibinfo{person}{Tat-Jun Chin}, \bibinfo{person}{Michael~S Brown}, {and} \bibinfo{person}{David Suter}.} \bibinfo{year}{2013}\natexlab{}.
\newblock \showarticletitle{As-projective-as-possible image stitching with moving DLT}. In \bibinfo{booktitle}{\emph{Proceedings of the IEEE conference on computer vision and pattern recognition}}. \bibinfo{pages}{2339--2346}.
\newblock


\bibitem[Zhang et~al\mbox{.}(2020)]%
        {zhang2020content}
\bibfield{author}{\bibinfo{person}{Jirong Zhang}, \bibinfo{person}{Chuan Wang}, \bibinfo{person}{Shuaicheng Liu}, \bibinfo{person}{Lanpeng Jia}, \bibinfo{person}{Nianjin Ye}, \bibinfo{person}{Jue Wang}, \bibinfo{person}{Ji Zhou}, {and} \bibinfo{person}{Jian Sun}.} \bibinfo{year}{2020}\natexlab{}.
\newblock \showarticletitle{Content-aware unsupervised deep homography estimation}. In \bibinfo{booktitle}{\emph{Computer Vision--ECCV 2020: 16th European Conference, Glasgow, UK, August 23--28, 2020, Proceedings, Part I 16}}. Springer, \bibinfo{pages}{653--669}.
\newblock


\bibitem[Zhang and Zhao(2024)]%
        {zhang2024mesa}
\bibfield{author}{\bibinfo{person}{Yesheng Zhang} {and} \bibinfo{person}{Xu Zhao}.} \bibinfo{year}{2024}\natexlab{}.
\newblock \showarticletitle{MESA: Matching Everything by Segmenting Anything}.
\newblock \bibinfo{journal}{\emph{arXiv preprint arXiv:2401.16741}} (\bibinfo{year}{2024}).
\newblock


\bibitem[Zhang et~al\mbox{.}(2023)]%
        {zhang2023searching}
\bibfield{author}{\bibinfo{person}{Yesheng Zhang}, \bibinfo{person}{Xu Zhao}, {and} \bibinfo{person}{Dahong Qian}.} \bibinfo{year}{2023}\natexlab{}.
\newblock \showarticletitle{Searching from Area to Point: A Hierarchical Framework for Semantic-Geometric Combined Feature Matching}.
\newblock \bibinfo{journal}{\emph{arXiv preprint arXiv:2305.00194}} (\bibinfo{year}{2023}).
\newblock


\bibitem[Zhang(2000)]%
        {zhang2000flexible}
\bibfield{author}{\bibinfo{person}{Zhengyou Zhang}.} \bibinfo{year}{2000}\natexlab{}.
\newblock \showarticletitle{A flexible new technique for camera calibration}.
\newblock \bibinfo{journal}{\emph{IEEE Transactions on pattern analysis and machine intelligence}} \bibinfo{volume}{22}, \bibinfo{number}{11} (\bibinfo{year}{2000}), \bibinfo{pages}{1330--1334}.
\newblock


\bibitem[Zhu and Liu(2023)]%
        {zhu2023pmatch}
\bibfield{author}{\bibinfo{person}{Shengjie Zhu} {and} \bibinfo{person}{Xiaoming Liu}.} \bibinfo{year}{2023}\natexlab{}.
\newblock \showarticletitle{Pmatch: Paired masked image modeling for dense geometric matching}. In \bibinfo{booktitle}{\emph{Proceedings of the IEEE/CVF Conference on Computer Vision and Pattern Recognition}}. \bibinfo{pages}{21909--21918}.
\newblock


\bibitem[Zou and Tan(2012)]%
        {zou2012coslam}
\bibfield{author}{\bibinfo{person}{Danping Zou} {and} \bibinfo{person}{Ping Tan}.} \bibinfo{year}{2012}\natexlab{}.
\newblock \showarticletitle{Coslam: Collaborative visual slam in dynamic environments}.
\newblock \bibinfo{journal}{\emph{IEEE transactions on pattern analysis and machine intelligence}} \bibinfo{volume}{35}, \bibinfo{number}{2} (\bibinfo{year}{2012}), \bibinfo{pages}{354--366}.
\newblock


\end{thebibliography}

%
\appendix

\section{Implementation Details}
\subsection{Architecture of Semantic-guide Interactions Block (SGIB)}
\begin{figure}[h]
\centering
\includegraphics[width=\linewidth]{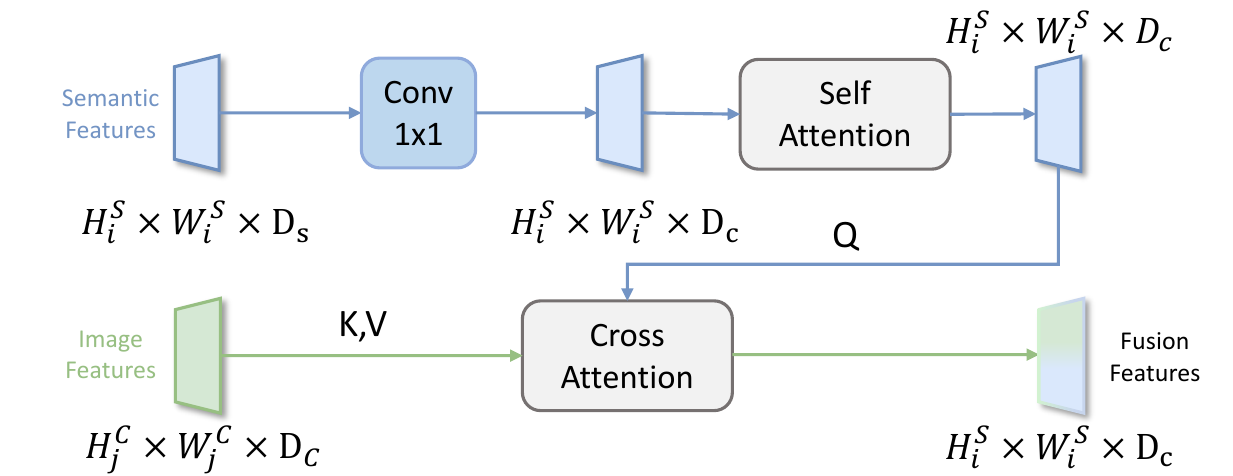}
\captionsetup{skip=5pt}
\caption{Architecture of Semantic-guide Interactions Block (SGIB). During stage 1, $i=j$; while in stage 2, $i\ne j$.}
\label{fig_s1}
\vspace{-8pt}
\end{figure}

In order to integrate image features and fine-grained semantic features, we propose the Semantic-guide Interactions Block (SGIB). The specific network architecture is shown in Figure \ref{fig_s1}. In stage 1, we first perform convolution layers to transform the semantic features to the same dimension as image features. The semantic features \(S_{i}\in \mathbb{R}^{H_{i}^{S} \times W_{i}^{S} \times D_{S}}\)$, i=0,1$ are going through the self-attention module and then fed into the cross-attention as query $Q$. The image features \(\hat{C_{j} }\in \mathbb{R}^{H_{j}^{C} \times W_{j}^{C} \times D_{C}}\)$, j=0,1$ are transmitted into the cross-attention treated as key $K$ and value $V$ to produce fusion features \(\tilde{C{} _{j}{}}{}'\in \mathbb{R}^{H_{i}^{S} \times W_{i}^{S} \times D_{C}}\), note that $H_{i}^{S}={H_{i}^{C}}, W_{i}^{S}={W_{i}^{C}}, i=0,1$. The calculation process of stage 2 is similar to stage 1.
\section{More Experiments}
\subsection{Performance of different methods}

Table \ref{tab:t1} shows the performance of various methods for homography estimation on HPatches \cite{balntas2017hpatches}. We utilize the MegaDepth \cite{li2018megadepth} dataset as the training dataset to retrain our SRMatcher following \cite{sun2021loftr}. We compare two groups of the methods, Detector-based and Detector-free methods.  Relative pose estimation results on the MegaDepth are given in Table \ref{tab:t2}. 

\begin{table}[h]
\caption{Evaluation on HPatches \cite{balntas2017hpatches} for homography estimation. For each method, a star symbol (*) denotes the official version trained on the Oxford-Paris dataset, whereas versions without a star indicate its official release trained on the author-preferred dataset. The \textbf{best} and \underline{second} results are highlighted.}\label{tab:PAUC_SN}
\centering
\resizebox{\linewidth}{!}{
\begin{tabular}{llllllll}
\toprule
\multicolumn{3}{l}{\multirow{3}{*}{Method}} & \multicolumn{4}{c}{Homography est. AUC$\uparrow$} \\ \cmidrule(l){4-7} 
\multicolumn{3}{c}{}                        & @3px       & @5px       & @10px   &mAUC     \\ \midrule
\multicolumn{6}{l}{$Detector$-$based$ $matching:$} \\

\multicolumn{3}{l}
{Superpoint~\cite{detone2018superpoint}~\tiny{CVPRW'18}}               & 43.4      & 57.6      & {72.7}   & 57.9  \\
\multicolumn{3}{l}
{SIFT~\cite{lowe2004distinctive}}               & 46.3   & 57.4      & {70.3} &58.0    \\
\multicolumn{3}{l}
{R2D2~\cite{revaud2019r2d2}~\tiny{NIPS'19}}               & 50.6      & 63.9      & {76.8}  &63.8   \\
\multicolumn{3}{l}
{SuperGlue~\cite{sarlin2020superglue}~\tiny{CVPR'20}}               & 53.9      & 68.3      & {81.7}  &68.0    \\
\midrule
\multicolumn{6}{l}{$Detector$-$free$ $matching:$} \\
\multicolumn{3}{l}{LoFTR*~\cite{sun2021loftr}~\tiny{CVPR'21}}  & 58.5  & 69.8 & 81.1 &69.8  \\
\multicolumn{3}{l}{LoFTR~\cite{sun2021loftr}~\tiny{CVPR'21}}  & 65.9  & 75.6 & 84.6 &75.4  \\
\multicolumn{3}{l}{QuadTree~\cite{tang2022quadtree}~\tiny{ICLR'22}}  & 66.3& 76.2 & 84.9 &75.8  \\
\multicolumn{3}{l}{ASpan~\cite{chen2022aspanformer}~\tiny{ECCV'22}}  & 67.4 & 76.9& 85.6 &76.6      \\
\multicolumn{3}{l}{TopicFM~\cite{giang2023topicfm}~\tiny{AAAI'23}}  & 67.3 & 77.0& 85.7&76.7       \\
\multicolumn{3}{l}{GeoFormer*~\cite{liu2023geometrized}~\tiny{ICCV'23}}                   & 68.0       & 76.8       & 85.4     &76.7  \\
\multicolumn{3}{l}{\textbf{SRMacther\_LoFTR}, trained on MegaDepth}  & 68.9  &76.9  & 84.9 &76.9  \\
\multicolumn{3}{l}{SEM~\cite{chang2023structured}~\tiny{CVPRW'23}}  & 69.6 & 79.0  & 87.1 &78.6      \\
\multicolumn{3}{l}{MESA~\cite{zhang2024mesa}~\tiny{CVPR'24}}      & {71.1}      & {{78.6}}      & {86.0}   &78.6    \\
\multicolumn{3}{l}{\textbf{SRMacther\_GeoFormer*}}      & {71.2}      & {{79.3}}      & {87.0}  &79.2     \\

\multicolumn{3}{l}{CasMTR-2c~\cite{cao2023improving}~\tiny{ICCV'23}}        & {71.4}      & {{80.2}}      & {87.9}  &79.8     \\ 
\multicolumn{3}{l}{GeoFormer, trained on MegaDepth~\cite{liu2023geometrized}~\tiny{ICCV'23}}                   & 72.1       & 79.9       & 87.7     &79.9  \\

\multicolumn{3}{l}{ASTR~\cite{yu2023astr}~\tiny{CVPR'23}}      & {71.7}      & {{80.3}}      & {88.0}   &80.0    \\
\multicolumn{3}{l}{DKM~\cite{edstedt2023dkm}~\tiny{CVPR'23}}      & {71.3}      & {{80.6}}      & \underline{88.5}   &80.1    \\
\multicolumn{3}{l}{PMatch~\cite{zhu2023pmatch}~\tiny{CVPR'23}}      & {71.9}      & {{80.7}}      & \underline{88.5}   &80.4    \\

\multicolumn{3}{l}{RoMa~\cite{edstedt2023roma}~\tiny{CVPR'24}}      & \underline{72.2}      & \underline{{81.2}}      & \textbf{89.1}  &\underline{80.8}     \\
\multicolumn{3}{l}{\textbf{SRMacther\_GeoFormer}, trained on MegaDepth}      & \textbf{73.5}      & \textbf{{81.3}}      & {88.0}   &\textbf{80.9}    \\
\bottomrule
\end{tabular}
}\label{tab:t1}
\end{table}

\begin{table}[h]
\caption{{Relative pose estimation results $(\%)$ on MegaDepth-1500 benchmark.} Training data: MegaDepth}\label{tab:PAUC_SN}
\centering
\resizebox{\linewidth}{!}{
\begin{tabular}{lllllll}
\toprule
\multicolumn{3}{l}{\multirow{3}{*}{Pose estimation AUC}} & \multicolumn{3}{c}{MegaDepth1500 benchmark} \\ \cmidrule(l){4-6} 
\multicolumn{3}{c}{}                        & AUC@5$^\circ\uparrow$       & AUC@10$^\circ\uparrow$        & AUC@20$^\circ\uparrow$        \\ \midrule
\multicolumn{3}{l}{TopicFM~\cite{giang2023topicfm}~\tiny{AAAI'23}}                   & 54.1       & 70.1       & 81.6       \\
\multicolumn{3}{l}{CasMTR-2c~\cite{cao2023improving}~\tiny{ICCV'23}}        & {59.1}      & {{74.3}}      & {84.8}       \\ 
\multicolumn{3}{l}{RoMa~\cite{edstedt2023roma}~\tiny{CVPR'24}}        & {62.6}      & {{76.7}}      & {86.3}       \\ 
\midrule
\multicolumn{3}{l}{LoFTR~\cite{sun2021loftr}~\tiny{CVPR'21}}                   & 52.8       & 69.2       & 81.2       \\
\multicolumn{3}{l}{SRMatcher\_LoFTR}        & {53.8}       & {70.4}       & {82.5}      \\
\multicolumn{3}{l}{GeoFormer~\cite{liu2023geometrized}~\tiny{ICCV'23}}                   & 51.7       & {68.3}       & {80.2}       \\
\multicolumn{3}{l}{SRMatcher\_GeoFormer}        & {53.2}       & {70.0}       & {81.8}      \\
\bottomrule
\end{tabular}
}\label{tab:t2}
\end{table}

\subsection{Ablations about fusion strategies}
Within SGIB, a variety of fusion techniques are available to incorporate priors, such as spatial attention and channel attention. We compare the cross-attention between semantic feature and image feature with these fusion strategies. As shown in Table \ref{tab:t3}, channel attention compromises the spatial integrity of semantics, resulting in the poorest performance. The performances of the spatial attention are also noticeably inferior to our proposed SGIB. This suggests that Semantic-Guide Interactions Block (SGIB) effectively implements cross-attention between semantic and image features, optimally utilizing the extensive semantic information available in Vision Foundation Models (VFMs) while maintaining spatial integrity.

\begin{table}[h]

\caption{Ablations about different fusion strategies in SFB.}
\resizebox{\linewidth}{!}{ 
\begin{tabular}{cccccccc}
\toprule
\multicolumn{3}{l}{\multirow{3}{*}{Modification}} & \multicolumn{4}{c}{Homography est. AUC} \\ \cmidrule(l){4-7} 
\multicolumn{3}{c}{}                        & @3px       & @5px        & @10px  & @mAUC     \\ \midrule
\multicolumn{3}{l}{Channel Attention} & 69.7     & 78.0       & 85.7          & 77.8   \\

\multicolumn{3}{l}{Spatial Attention}        & 70.1   & 78.5      & 86.0 & 78.2     \\
\multicolumn{3}{l}{SRMatcher\_GeoFormer }           & \textbf{71.2}   &\textbf{79.3}       & \textbf{87.0}  & \textbf{79.2}   \\
\bottomrule
\end{tabular}
}\label{tab:t3}
\vspace{-0.8em} 
\end{table}
\subsection{Ablations about Semantic Extractors}
To investigate whether fine-grained semantic features enhance the efficacy of matching results, we employ three distinct semantic extractors. The first one is ResNet-50 \cite{he2016deep} pre-trained on the ImageNet. Another one is the pre-trained CLIP \cite{radford2021learning} which has a strong semantic extraction ability due to its text-image pairs training method. As Table \ref{tab:t4} shows, the features generated by DINOv2 \cite{oquab2023dinov2} are more effective than the others. This superior performance stems from DINOv2's self-supervised training method, which compels the model to learn image features that are consistent across various transformations and inherently possess a high semantic value.  

\begin{table}[h]

\caption{Ablations about different semantic extractors. 'R', 'C', 'D' denote the ResNet-50, CLIP and DINOv2.} \label{tab:ASR}
\resizebox{\linewidth}{!}{ 
\begin{tabular}{cccccccc}
\toprule
\multicolumn{3}{l}{\multirow{3}{*}{Modification}} & \multicolumn{4}{c}{Homography est. AUC} \\ \cmidrule(l){4-7} 
\multicolumn{3}{c}{}                        & @3px       & @5px        & @10px  & @mAUC     \\ \midrule
\multicolumn{3}{l}{SRMatcher-R}      & 69.7       & 78.3    & 86.2  &78.0  \\

\multicolumn{3}{l}{SRMacther-C}        &70.5   & 78.5      & 86.6 & 78.5     \\
\multicolumn{3}{l}{SRMatcher-D}           & \textbf{71.2}   &\textbf{79.3}       & \textbf{87.0}  & \textbf{79.2}   \\
\bottomrule
\end{tabular}
}\label{tab:t4}
\vspace{-0.8em} 
\end{table}
\subsection{Ablations about Different Layers}
It is important to note that the semantic content extracted from different layers of DINOv2 varies. As the number of layers increases, the semantics become more representative. The key to successful semantic guidance is to extract semantic features that are both deep and capable of retaining critical spatial information, which is vital for ensuring effective semantic guidance. In our initial experimental design, following previous methods \cite{li2024sed, liu2024visual} that utilized vision foundation models (VFMs), we opted not to use features from the last layer of VFMs. This issue was made because vision foundation models (VFMs), being pretrained for specific downstream tasks, may not be well-suited for task transfer. Therefore, in the main text, we use features from the third-to-last layer of DINOv2 as semantic priors. However, as shown in Tabel \ref{tab:t5} we tried using the features from different layers as the semantics. We find that the the features from last layer lead to the performance increase, this finding differs from previous methods. We believe this is due to DINOv2 being pretrained through image-level and patch-level discriminative self-supervised learning, which enables it to extract all-purpose visual features, thus facilitating zero-shot patch-level feature matching capability \cite{liu2023matcher}. 

\begin{table}[h]

\caption{Ablations about different layers of DINOv2.}
\resizebox{\linewidth}{!}{ 
\begin{tabular}{cccccccc}
\toprule
\multicolumn{3}{l}{\multirow{3}{*}{Modification}} & \multicolumn{4}{c}{Homography est. AUC} \\ \cmidrule(l){4-7} 
\multicolumn{3}{c}{}                        & @3px       & @5px        & @10px  & @mAUC     \\ \midrule

\multicolumn{3}{l}{Third to last \textbf{(main text)}}    & {71.2}   &{79.3}  & {87.0}  & {79.2}   \\
\multicolumn{3}{l}{Second to last}        &71.4   & 79.7    & 87.4   & 79.5     \\
\multicolumn{3}{l}{Last}    & \textbf{71.8}    & \textbf{79.9}  & \textbf{87.6}  &\textbf{79.8}  \\
\bottomrule
\end{tabular}
}\label{tab:t5}
\vspace{-0.8em} 
\end{table}
\subsection{Ablations about Semantic Extractor parameter}
The DINOv2 is employed as semantic extractor to obtain various and available semantic information. Specifically, we use DINOv2 with a ViT-B/14 \cite{dosovitskiy2020image} with registers as the default semantic extractor of SRMatcher. We also conduct comparison experiments on ViT-S/14 and ViT-L/14 with different parameters shown in Table \ref{tab:t6}.
\begin{table}[h]
\centering
\caption{Ablations about DINOv2 parameters.} \label{tab:t6}
\resizebox{\linewidth}{!}{ 
\begin{tabular}{cccc|ccccc}  
\toprule
\multicolumn{4}{c}{\multirow{2}{*}{Modification}} & \multirow{2}{*}{Params} & \multicolumn{4}{c}{Homography est. AUC} \\ 
\cmidrule{6-9}  
\multicolumn{4}{c}{} & & @3px & @5px & @10px & mAUC \\  
\midrule
\multicolumn{4}{l}{DINOv2\_ViT-S/14 } & 21M & 70.5 & 78.9 & 86.5 & 78.6 \\  
\multicolumn{4}{l}{DINOv2\_ViT-B/14 } & 86M  & {71.2} & {79.3} & {87.0} & {79.2} \\  
\multicolumn{4}{l}{DINOv2\_ViT-L/14 } & 300M & \textbf{71.3} & \textbf{79.4} & \textbf{87.3} & \textbf{79.3} \\  
\bottomrule
\end{tabular}
}
\vspace{-0.6em}
\end{table}

\subsection{Ablations about Stage Repetition in SFB}
We repeat stage 2 as ablation study. Experiments show adding more stages indeed brings improvement, but it also leads to increased computational costs. Users can select based on their requirements. In this paper, we choose the two stages for balancing performance and computation as shown in Table \ref{tab:t7}.
\begin{table}[htbp]
\centering
\caption{Ablations about Stage Repetition in SFB.} \label{tab:t7}
\resizebox{0.46\textwidth}{!}{
\begin{tabular}{ccccccccc}
\hline
\multirow{2}{*}{est. AUC} & \multicolumn{4}{c}{Stage Repetition}  \\ \cline{2-5} 
                          & s1   & s1+1$\times$s2 & s1+2$\times$s2 & s1+3$\times$s2      \\ \hline
3px                       & 69.8 & 71.2 & \textbf{71.8} & 71.6   \\ 
mAUC                       & 77.7 & 79.2 & 79.5  & \textbf{79.9}  \\
training params/M             & 17.1 & 19.7 & 22.4 & 25.0     \\      
 \hline
\end{tabular}
\label{tab:1}
}
\end{table}

\subsection{Ablations about Tuning Methods}
For the fine-tuning methods for pre-traing model, we employed three fine-tuning methods: tuning the last three layers, its EMA, and LoRA. The experiments show the result that the model achieved superior performance after fine-tuning as shown in Table \ref{tab:t8}.

\begin{table}[htbp]
\centering
\caption{Ablations about Tuning Methods.} \label{tab:t8}
\resizebox{0.46\textwidth}{!}{
\begin{tabular}{ccccccccc}
\hline
\multirow{2}{*}{est. AUC} &  \multicolumn{4}{c}{Tuning Methods} \\ \cline{2-5} 
                           &frozen  & last three   & EMA & LoRA     \\ \hline
3px                        &71.2& \textbf{71.8} & 71.6 & \textbf{71.8}  \\ 
mAUC                       &79.2& 79.7 & 79.4 & \textbf{79.9}  \\
      
 \hline
\end{tabular}
\label{tab:1}
}

\end{table}
\subsection{Ablations about Different Scales}
When we extend our model with a more coarse level (1/16), SRMatcher cannot achieve proper matching results, matching will cause inevitable errors for subsequent learning as shown in Table \ref{tab:t9}. Less coarse level (1/4) causes Out-of-memory (OOM) errors in a 24GB GPU due to their long sequences.

\begin{table}[htbp]
\caption{Ablations about Different Scales.}
\centering
\resizebox{0.46\textwidth}{!}{
\begin{tabular}{ccccccccc}
\hline
\multirow{2}{*}{est. AUC} & \multicolumn{5}{c}{Different Scales}  \\ \cline{2-6}
                            & 1/16+1/4 & 1/16+1/2 & 1/8+1/4 & 1/8+1/2 & 1/4+1/2        \\ \hline
3px                      & 63.4 & 64.7 & 70.6 & \textbf{71.2} &OOM  \\ 
mAUC                      & 72.8 & 74.0 & 78.4 & \textbf{79.2} &OOM \\ 
 \hline
\end{tabular}
\label{tab:t9}
}

\end{table}

\section{Generalizability for Realistic Scenarios}
Homography is typically assumed to be a rigid transformation, but in reality, some scenes may involve non-rigid transformations. Our SRMatcher is equipped to dealing with scenarios that contain non-rigid transformations. For instance, in motion capture tasks, it is necessary to handle the non-rigid movement of body parts. Feature matching method can be used to locate specific body keypoints. Using these keypoints, a Thin Plate Spline (TPS) motion model is constructed to reconstruct the movement. By incorporating semantic information, SRMatcher can accurately identify correspondence under non-rigid transformations by recognizing objects whose shapes change but whose semantics remain consistent. 

\section{Matching in Extreme Cases}
Semantic extractors used in SRMatcher are derived from vision foundation models pre-trained on large-scale datasets. When dealing with some cases such as occluded scenes, extreme lighting variations, or low contrast images, these models \cite{amir2021deep, kirillov2023segment, radford2021learning, oquab2023dinov2} have been demonstrated to effectively \textbf{extract rich semantic features} across various complex scenarios, fusion with the low-level image features. We propose the \textbf{cross-image fusion strategy}, which not only considers the semantic information within a single image but also integrates semantics from a paired image. When not enough information can be extracted, this strategy utilizes information from the other image to enhance deficient regions in the current view. As shwon in Table \ref{tab:t10}, experiments conducted on the extreme scenarios dataset such as extreme lighting changes (selected in HPatches, its performance denotes as mAUC\_E), show that substituting different semantic extractors still yields satisfactory performance and the effectiveness of the cross-image fusion strategy.
\begin{table}[htbp]

\centering
\caption{Experiments conducted on the extreme scenarios dataset}
\resizebox{0.47\textwidth}{!}{
\begin{tabular}{cccccccccc}
\hline
est. AUC &Baseline& CLIP   & DINOv1  & SAM & DINOv2  &w/o Cross-image fusion    \\ \hline
 
 mAUC\_E& 65.9 &73.2& 73.4&73.7 & \textbf{74.3} & 73.1 \\ 
 \hline
\end{tabular}
\label{tab:t10}
}

\end{table}
\section{Qualitative Results}
\begin{figure*}
  \includegraphics[width=\textwidth]{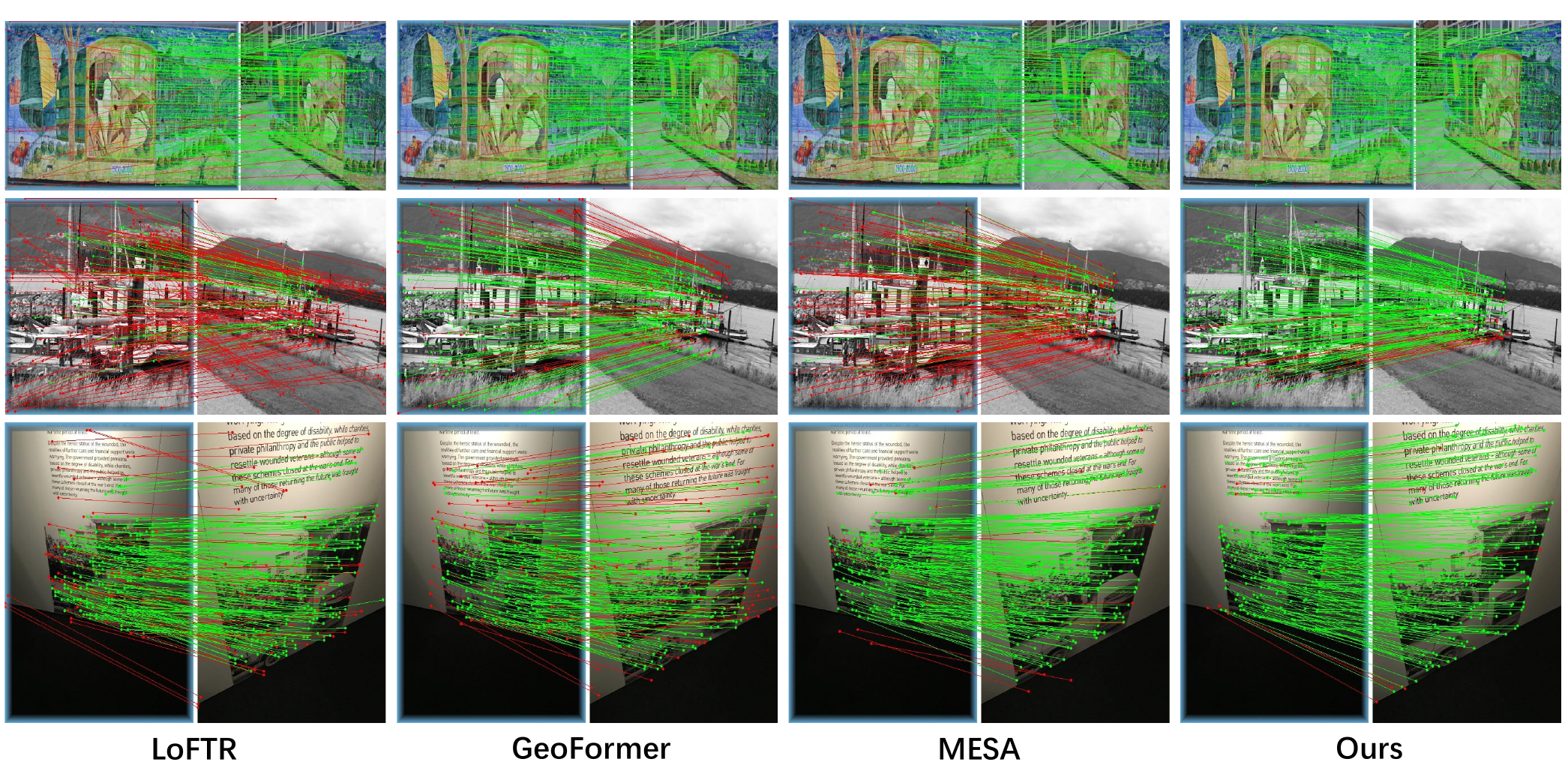}
  \captionsetup{skip=0.2cm}
  \caption{Qualitative of matching results with LoFTR \cite{sun2021loftr}, GeoFormer \cite{liu2023geometrized}, MESA \cite{zhang2024mesa}, and our SRMatcher on HPatches \cite{balntas2017hpatches}. Points classified as inliers by RANSAC are displayed in green, while outliers are shown in red. }
  \label{fig_2}
  \vspace{0.5cm}
\end{figure*}

\begin{figure*}
  \includegraphics[width=\textwidth]{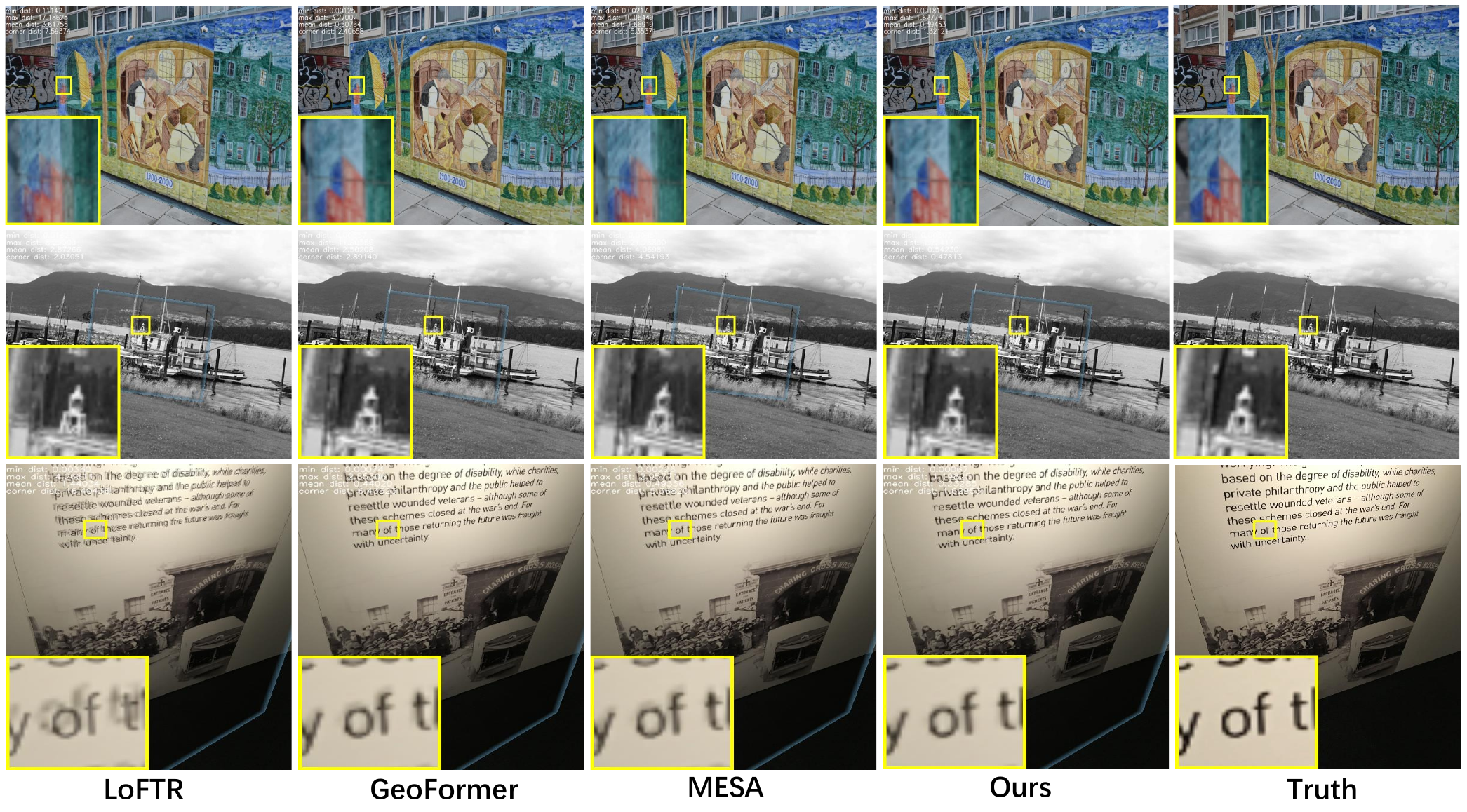}
  \captionsetup{skip=0.2cm}
  \caption{Qualitative of homography estimation results with LoFTR \cite{sun2021loftr}, GeoFormer \cite{liu2023geometrized}, MESA \cite{zhang2024mesa}, and our SRMatcher on HPatches \cite{balntas2017hpatches}. }
  \label{fig_3}
\end{figure*}

\begin{figure*}
  \includegraphics[width=\textwidth]{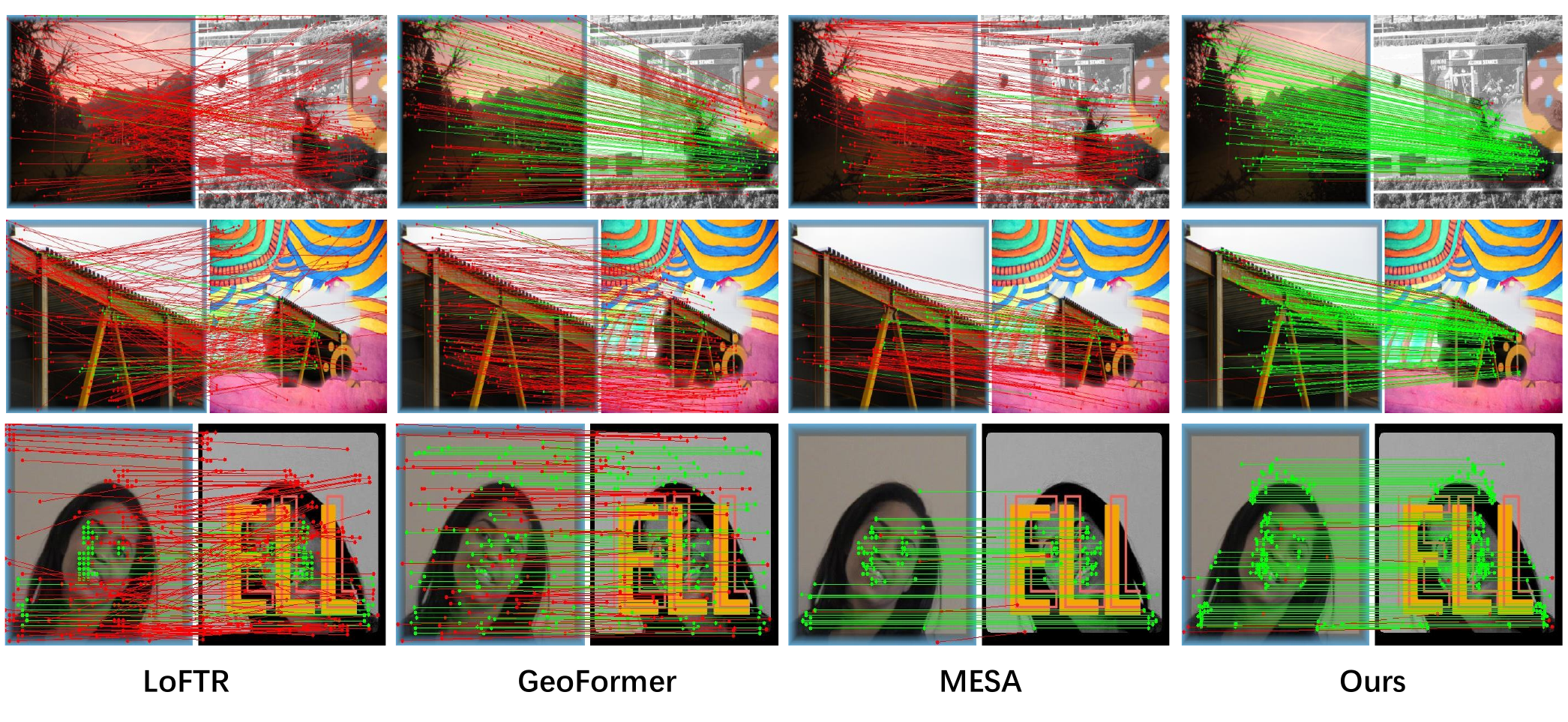}
  \captionsetup{skip=0.2cm}
  \caption{Qualitative of matching results with LoFTR \cite{sun2021loftr}, GeoFormer \cite{liu2023geometrized}, MESA \cite{zhang2024mesa}, and our SRMatcher on ISC-HE \cite{liu2023geometrized}. Points classified as inliers by RANSAC are displayed in green, while outliers are shown in red. }
  \label{fig_4}
  \vspace{0.5cm}
\end{figure*}

\begin{figure*}
  \includegraphics[width=\textwidth]{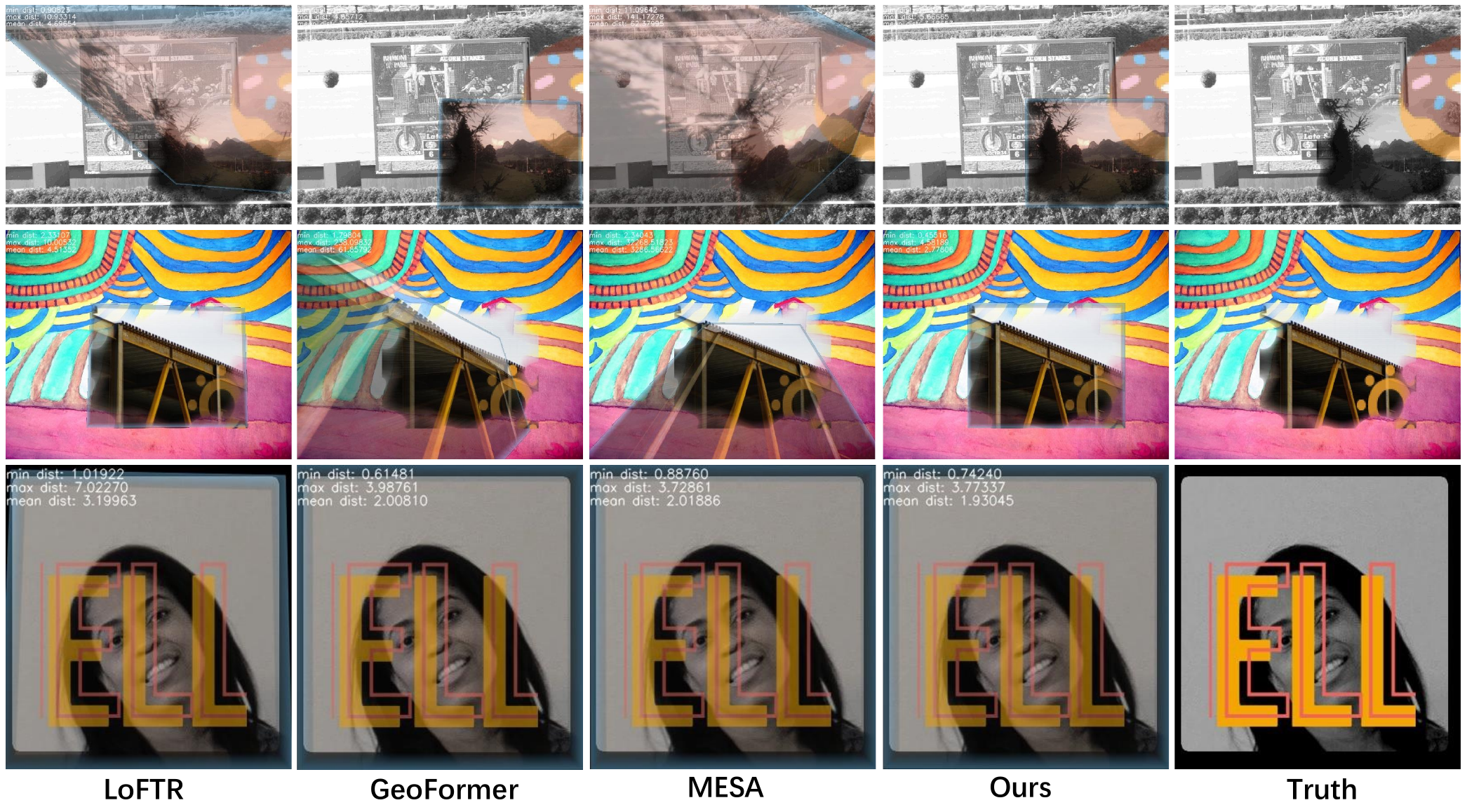}
  \captionsetup{skip=0.3cm}
  \caption{Qualitative of homography estimation results with LoFTR \cite{sun2021loftr}, GeoFormer \cite{liu2023geometrized}, MESA \cite{zhang2024mesa}, and our SRMatcher on ISC-HE \cite{liu2023geometrized}. }
  \label{fig_5}
\end{figure*}

\begin{figure*}
  \includegraphics[width=\textwidth]{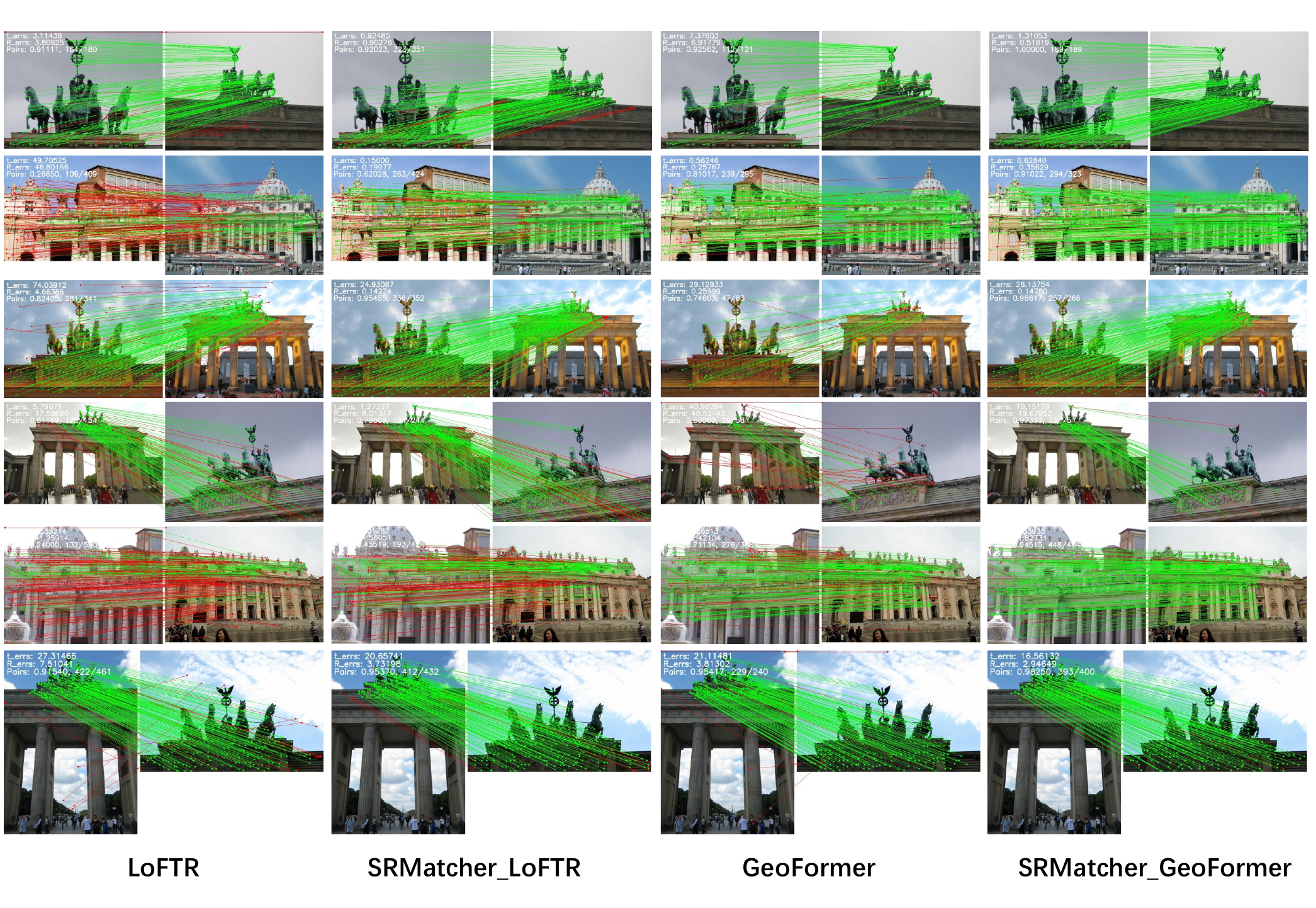}
  \captionsetup{skip=0.2cm}
  \caption{Qualitative image matches on MegaDepth dataset. Green signifies that the epipolar error in normalized image coordinates is below $1\times 10^{-4} $, whereas red denotes that this threshold has been surpassed. Training data: MegaDepth.}
  \label{fig_6}
  \vspace{0.5cm}
\end{figure*}

We provide additional qualitative comparisons of SRMatcher and baseline methods on the Hpatches \cite{balntas2017hpatches}, ISC-HE \cite{liu2023geometrized} and MegaDepth \cite{li2018megadepth} datasets. In Figure \ref{fig_2} and Figure \ref{fig_4}, we illustrate inlier and outlier matches with various projection thresholds to evaluate the matching precision of different methods on the Hpatches dataset and ISC-HE dataset. Figure \ref{fig_3} and Figure \ref{fig_5} display further qualitative results of homography estimation, the methods being compared include LoFTR \cite{sun2021loftr}, GeoFormer \cite{liu2023geometrized}, MESA \cite{zhang2024mesa}, and our SRMatcher\_GeoFormer.
Figure \ref{fig_6} offers more qualitative insights on the MegaDepth dataset, the methods being compared include LoFTR, SRMatcher\_LoFTR, GeoFormer, SRMactcher\_GeoFormer trained on MegaDepth dataset.
\clearpage

\clearpage

\clearpage

\end{document}